\documentclass[superscriptaddress]{article}
\usepackage{authblk}
\usepackage{graphics}
\usepackage[dvips]{graphicx}
\usepackage{subfig}
\usepackage{amsmath}
\usepackage{hyperref}

\date{\today}

\begin{document}

\title{Evolutionary dynamics of higher-order interactions in social networks}

\author[1,2,*]{Unai Alvarez-Rodriguez}
\author[3,4]{Federico Battiston}
\author[5]{Guilherme Ferraz de Arruda}
\author[5,6,7]{Yamir Moreno}
\author[8,9,10]{Matja{\v z} Perc}
\author[2,11,12]{Vito Latora}

\affil[1]{Basque Centre for Climate Change (BC3), 48940, Leioa, Spain}
\affil[2]{School of Mathematical Sciences, Queen Mary University of London, London E1 4NS, UK}
\affil[3]{Department of Network and Data Science, Central European University, Vienna, Austria}
\affil[4]{Department of Anthropology, University of Zurich, Zurich, Switzerland}
\affil[5]{ISI Foundation, Turin, Italy}
\affil[6]{Institute for Biocomputation and Physics of Complex Systems, University of Zaragoza, Zaragoza 50008, Spain}
\affil[7]{Department of Theoretical Physics, University of Zaragoza, Zaragoza 50009, Spain}
\affil[8]{Faculty of Natural Sciences and Mathematics, University of Maribor, Koro{\v s}ka cesta 160, 2000 Maribor, Slovenia}
\affil[9]{Department of Medical Research, China Medical University Hospital, China Medical University, Taichung, Taiwan}
\affil[10]{Complexity Science Hub Vienna, Josefst{\"a}dterstra{\ss}e 39, 1080 Vienna, Austria}
\affil[11]{Dipartimento di Fisica ed Astronomia, Universit{\`a} di Catania and INFN, 95123 Catania, Italy}
\affil[12]{The Alan Turing Institute, The British Library, London, NW1 2DB, UK}

\affil[*]{Corresponding author: Unai Alvarez-Rodriguez (unaialvarezr@gmail.com; ORCID linked to account on Manuscript Tracking System)}

\maketitle

\clearpage

\subsection*{Abstract}
\textbf{We live and cooperate in networks. However, links in networks only allow for pairwise interactions, thus making the framework suitable for dyadic games, but not for games that are played in groups of more than two players. Here, we study the evolutionary dynamics of a public goods game in social systems with higher-order interactions. First, we show that the game on uniform hypergraphs corresponds to the replicator dynamics in the well-mixed limit, providing a formal theoretical foundation to study cooperation in networked groups. Secondly, we unveil how the presence of hubs and the coexistence of interactions in groups of different sizes affects the evolution of cooperation. Finally, we apply the proposed framework to extract the actual dependence of the synergy factor on the size of a group from real-world collaboration data in science and technology. Our work provides a way to implement informed actions to boost cooperation in social groups.}

\section*{Introduction}
Cooperation among unrelated individuals distinguishes humans markedly from other mammals, and it is one of the central pillars of our evolutionary success \cite{nowak_11}. Past research has emphasized that the structure of social interactions is crucial for the evolution of cooperation, but thus far predominantly in the realm of networks where links connect pairs of players \cite{santos_prsb06, rand_pnas11}. However, since cooperation often unfolds in groups, the need for a paradigm shift in the way we model social interactions is evident and indeed urgent. Regardless of the model that we use to describe human interactions, cooperation remains at odds with the fundamental principles of Darwinian evolution, and it is fascinating that we have succeeded in collectively holding off self-interest over most of the last two million years, ever since the genus Homo first emerged \cite{hrdy_11}.

Given this puzzle, the search for reasons and mechanisms that may
allow cooperation to evolve and proliferate is an evergreen and
vibrant subject across the social and natural sciences
\cite{henrich_aer01, nowak_s06, henrich2007humans, rand_tcs13,kraft_cobs15, perc_pr17,jackson13}.
Evolutionary game theory is long established as the theory of choice for addressing the puzzle
mathematically \cite{weibull_95, hofbauer_98, nowak_06}, wherein
social dilemmas constitute a particularly important class of
games. Namely, social dilemmas capture the essence of the problem
since defection is the individually optimal strategy, whilst
cooperation is the optimal strategy for the highest social welfare
\cite{axelrod_84}. An important mechanism for cooperation in social
dilemmas is network reciprocity \cite{nowak_n92b}, which stands for
the fact that a limited interaction range, as dictated by lattices or
other types of networks, facilitates the formation of compact clusters
of cooperators that are in this way protected against invading
defectors. This basic mechanism could also be seen if the degree
distribution of the interaction network is strongly heterogeneous
\cite{santos_prl05, santos_pnas06, gomez-gardenes_prl07}, if there is
set or community structure \cite{tarnita_pnas09, fotouhi_rsif19}, or
if the evolution unfolds on two or more network layers that mutually
support cooperative clusters \cite{wang_z_epl12,
  gomez-gardenes_srep12, gomez-gardenes_pre12, wang_z_srep13,
  wang_pre14a, battiston_njp17, fu2017leveraging,
  fotouhi2018conjoining}.

Despite the wealth of important insights concerning the evolution of cooperation on networks and fundamental discoveries \cite{lieberman_n05, ohtsuki_n06, allen2017evolutionary}, an important unsolved problem remains accounting for cooperation in groups, such as for example in the public goods game (PGG) \cite{archetti_jtb12, perc_jrsi13}. The simplest remedy is to consider members of a group to be all the players that are pairwise-connected to a central player \cite{santos_n08, szolnoki_pre09c}. However, since the other players are further connected in a pairwise manner, one would also need to consider all the groups in which the central player is a member but is not central. Evidently, classical networks do not provide a unique procedure for defining a group. Moreover, members of the same group are commonly not all directly connected with one another, which prevents strategy changes among them, either in terms of imitation, replication, or exploration. These facts posit a lack of common theoretical foundation for studying the evolution of cooperation in networked groups. Without knowing who is connected to whom in a group, it is also impossible to implement fundamental mechanisms that promote cooperation, such as reciprocity \cite{trivers_qrb71, sigmund_tee07}, image scoring \cite{nowak_n98, milinski_prslb01, nax_srep15b}, and reputation \cite{fehr_n04, gaechter_sje02, fu_pre08b}.

As a solution, we here introduce and study higher-order interactions
in evolutionary games that are played in groups. The distinctive
feature of higher-order interactions is that, unlike in classical
networks \cite{latora2017complex}, a link can connect more than just
two individuals \cite{berge1984hypergraphs}. Thus, higher-order
networks naturally account for structured group interactions \cite{fede20}, wherein
a group is simply made up of all players that are connected by a so-called hyperlink, which is the higher-order analogous of the link. As
a paradigmatic example, we consider a standard public goods game on
the higher-order analogous of a network, referred to as a hypergraph,
see Figure \ref{fig:fig1}. We first show that it corresponds
exactly to the replicator dynamics in the well-mixed limit as long as no hyperdegree-hyperdegree correlations exist. As such,
it thus provides a formal theoretical foundation to study cooperation
in networked groups -- effectively a null model -- that is amenable to
further upgrades. Next, we consider the public goods game on hypergraphs with heterogeneity either in their node hyperdegrees (number of hyperlinks a node is involved into) or in the order of their hyperlinks (number of nodes that form each hyperlink), which allow us to describe the dynamics induced by the presence of highly connected players and to consider scenarios in
which the synergy factor depends on the group size in a systematic and
consistent way. We show, for example, how synergy factors that are
given by different powers of the group size lead to a critical scaling
in the transition from defection to cooperation. Lastly, we also
demonstrate how the proposed higher-order interaction framework can be
used to determine the synergy factor as a function of the group size
from empirical data on cooperation and collaborations.
Under the assumption that the structure of the
hypergraph is the outcome of an optimisation process of the game it
hosts, we extract the game parameters from datasets describing collaborations in science and technology, showing that higher-order interactions induce diverse benefits and costs in different social domains.

The public goods game constitutes the fundamental
  example of a social dilemma when multiple individuals
  interact simultaneously. It presents a situation where the gain or
  loss of an initial investment is shared symmetrically between the
  members of a group, even if the investment itself can be
  asymmetric. In other words, there is no correlation between the
  individual effort and the distribution of the reward, meaning that
  some players receive more than what they give or deserve,
  while some others receive less.  
  Metaphorically, one would say that the game has no memory,
  in the sense that the payoff is assigned blindly to all the players
  as if the system had lost the information about the original
  contribution of each player. More formally, the public goods game
describes a setting where $N$ players are requested to contribute to a
common pool with a token of value $c$ \cite{perc_pr17}.

Cooperators do contribute, and defectors do not. The
collected amount is then multiplied by the so-called synergy factor
$R$, and the benefit is shared amongst all the members of the
group. The payoff for the defectors and cooperators playing in a group
of $g$ members is given by $\pi_D=R c w_C / g$ and $\pi_C= R c w_C /g
- c$ respectively, with $w_C$ representing the number of
cooperators in the group. Typically $c$ has a fixed value of $c=1$, so that the
behaviour of the system is determined by the synergy factor $R$, or
the reduced synergy factor $r=R/g$. Besides, it is common to represent
the state of the system by the fraction of players adopting each
strategy, $x_C$ for the cooperators and $x_D$ for the
defectors. 

The evolutionary dynamics determines how the strategies of the players
evolve with each iteration of the PGG, that is, how the fractions
$x_C$ and $x_D$ change with time. Here, we implement the so-called
fixed cost per game approach, where cooperators
contribute with an entire token to each game they play.  Individual
updates constitute micro-steps of the dynamics, whereas a (global)
time step corresponds to $N$ individual steps, so that all the players
in the system have the chance to play the game and update their
strategies. Players interact among them following the links of the
network they are embedded in. As mentioned before, the standard
network implementation \cite{santos_n08}, henceforth referred to as
graph implementation (GI), is not able to account
for the most general type of interaction in groups.
One of the first proposals to overcome the limitation of a GI is Evolutionary Set Theory \cite{tarnita_pnas09}, that considers a structure of interaction in which the players are organised as the elements of a set. Yet, the game itself is pairwise, and thus different from the type of approach proposed here. However, it is worth pointing out that the set theory description is equivalent to the hypergraph formalism, and therefore, one should expect the same results when studying the same game on both structures. In this work we have opted for hypergraphs because, as a higher-order generalization of graphs, they inherit the whole family of graph tools with which evolutionary game theory scholars are more familiar with. A few years later, it was proposed to address higher-order interactions by bipartite graphs, having a set of nodes for the players and a second set for the groups \cite{gomez-gardenes_c11,gomez-gardenes_epl11,pena_pone12}. The authors adapted the PGG to the bipartite graph, in what we call the bipartite implementation (BI). In such a case the game is indeed polyadic, but the update process is still dyadic, and the constrains associated to the formalism do not make it suitable for an analytical treatment. Here, we generalise the BI to a fully higher-order implementation and provide the theoretical foundation to study higher-order cooperative games in uniform and heterogeneous hypergraphs. Finally, we mention that in a more recent work \cite{wu19}, the authors have considered games played by agents belonging to subpopulations and whose interactions occur across and within the population, providing a useful methodology for situations in which one can get rid of the fine details of the individual connections.

\section*{Results}

\subsection*{Game Implementation}
\begin{figure}[h!]
\includegraphics[width=\textwidth]{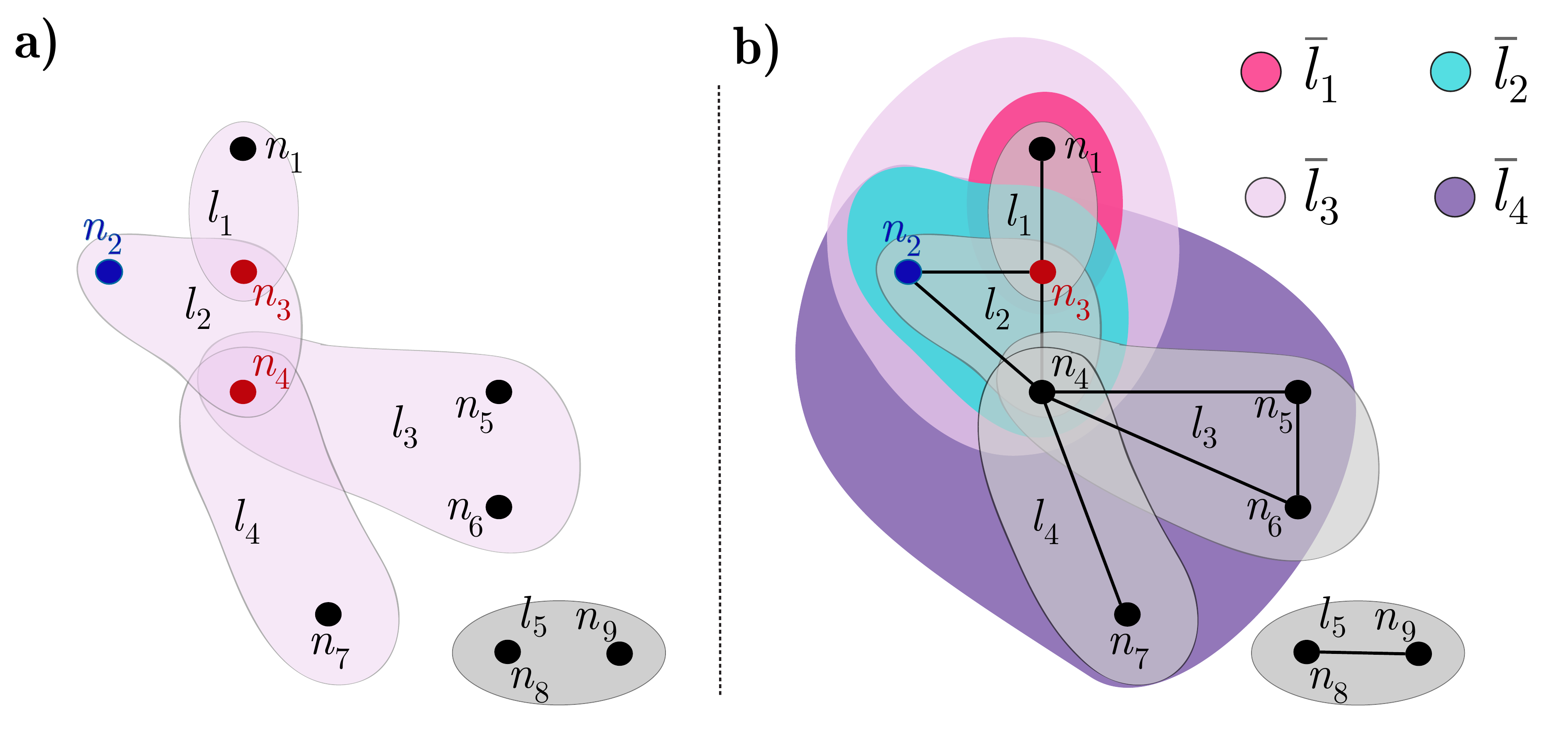}
\caption{Higher-order vs pairwise interactions in a Public Goods Game (PGG). Comparison of the proposed hypergraph implementation (HI) with a standard graph implementation (GI) of the game based on pairwise interactions only. (a) In the HI implementation, a node, $n_2$, and one of its hyperlinks, $l_2$, are randomly selected. All the nodes in $l_2$, namely node $n_2$, and the two nodes highlighted in red $n_3$ and $n_4$, play all the games they are involved in, corresponding, in this example, to PGG defined for the subset of nodes of the hyperlinks $l_1$, $l_2$, $l_3$ and $l_4$. Then, the strategy of $n_2$ is updated by comparing its payoff with that of the node with the highest accumulated payoff of the hyperlink $l_2$. This is not equivalent to play the PGG in the graph generated by projecting the interactions of the hypergraph, which is shown in (b). In the standard GI implementation, a neighbour of $n_2$, let us say $n_3$ $-$highlighted in red$-$ is randomly selected. The two nodes $n_2$ and $n_3$ then play all the games of the groups they are part of, that is, of the groups made up by the subsets of nodes $\{n_1,n_3\}$, $\{n_2,n_3,n_4\}$, $\{n_1,n_2,n_3,n_4\}$ and $\{n_2,n_3,n_4,n_5,n_6,n_7\}$. These subsets, coloured as indicated in the figure, could be represented by a different set of hyperlinks $\bar{l}_1$, $\bar{l}_2$, $\bar{l}_3$ and $\bar{l}_4$, respectively, which are different from the set of hyperlinks of the original hypergraph. Finally, the strategy of $n_2$ is updated by comparing its accumulated payoff to that of node $n_3$.}
\label{fig:fig1}
\end{figure}

In order to account for higher-order interactions, we use hypergraphs \cite{berge1984hypergraphs}.
A hypergraph, $H(\mathcal{N},\mathcal{L})$, is a mathematical object that consists of a set of $N$ nodes $\mathcal{N}=\{n_1=1,..,n_N=N\}$ and a set of $L$ hyperlinks $\mathcal{L}=\{l_1,...,l_L\}$. Each hyperlink is a subset of two or more elements of $\mathcal{N}$ and represents a group interaction. For instance, in Figure \ref{fig:fig1}a, the hyperlink $l_1$ contains nodes $n_1$ and $n_3$, whereas the hyperlink $l_3$ is the subset made up by nodes $n_4$, $n_5$ and $n_6$. Furthermore, the cardinality of a subset, known as the order of the hyperlink, is the number $g$ of nodes in the group. In the previous example, $l_1$ has order 2 and $l_3$ has order 3. In a hypergraph, the hyperdegree, $k_i$, of a node $i$ represents the number of hyperlinks in which the node is involved into, thus, the number of groups of a specific order $g$ that contains $i$ can be denoted by $k^g_i$. Hence, the hyperdegree of $i$ is given as $k_i = \sum^{g^+}_{g=g^-} k_i^g$, where $g^-$ and $g^+$ account for the minimal and maximal orders in $\mathcal{L}$. For example, in Figure \ref{fig:fig1}a, $k_4=3=k^2_4+k^3_4$, with $k^2_4=1$ (the hyperlink $l_4$) and $k^3_4=2$ (the hyperlinks $l_2$ and $l_3$).
As $\langle k \rangle$ we indicate the average hyperdegree of node $i$, where the averages are evaluated over all the nodes in the system, i.e. $ \langle  k \rangle = \frac{1}{N} \sum_{i\in \mathcal{N}} k_i$.

Although hypergraphs are not the only possible representation of group interactions, they allow exploiting the analogy between the links representing pairwise interactions in contact networks and hyperlinks, which are based on higher-order, group interactions. As we will show next, the differences between these two approaches lead to fundamentally distinct outcomes of the PGG evolutionary dynamics. To see how the evolutionary dynamics evolves in hypergraphs, let us consider the first step of a standard graph implementation of the PGG. When a node $n_i$ and one of its neighbours $n_j$ are selected on a graph, it is equivalent to say that a node and one of its links are selected. Such a procedure can be easily generalised to group interactions of more than $g=2$ individuals, see Figure \ref{fig:fig1}b. Note that if we choose more neighbours of $n_i$ to generate higher-order interactions, such an extension would still be based on dyadic ones. Instead, we propose a hypergraph implementation (HI) of the game that consists of selecting one of the hyperlinks of $n_i$. That is, in the HI setup, we select at random with uniform probability a node $n_i$ in the hypergraph and one of its hyperlinks, $l_i$. Then, all the members of the hyperlink $l_i$ play a game for each of the hyperlinks they are part of, as illustrated in Figure \ref{fig:fig1}. Finally, as it is customary, the nodes accumulate the payoffs of all the rounds they play, and we normalise this quantity by the total number of played games, such that each node's performance is represented by its average payoff.

The second part of each micro-step of the evolutionary dynamics of the game involves updating the strategy of node $n_i$. To this end, we normalise the discrete replicator dynamics for the case of higher-order interactions. We propose to compare the payoff $\pi_i$ of a node $n_i$ with the maximal payoff of the selected hyperlink $l_i$. Under this rule, $n_i$ will adopt the strategy of the node with the maximal payoff with a probability $\frac{1}{\Delta} [(\max_{l_i} \pi_j) -\pi_i]$, where $\Delta$, whose precise definition is provided in Equation \eqref{delta}, accounts for the maximal payoff difference, and is employed to guarantee that the probability is normalised. The rationale behind the choice of this expression is that node $i$ will compare its payoff to that of the node with the largest payoff in hyperlink $l_i$. Note that the previous expression reduces to the standard one of the $GI$ when $g=2$. Summing up, the HI accounts for a more realistic update than that in the BI, since the player inspiring a strategy change is the one with the highest payoff of the group, and not a randomly chosen one.

\begin{figure}[h]
\includegraphics[width=\textwidth]{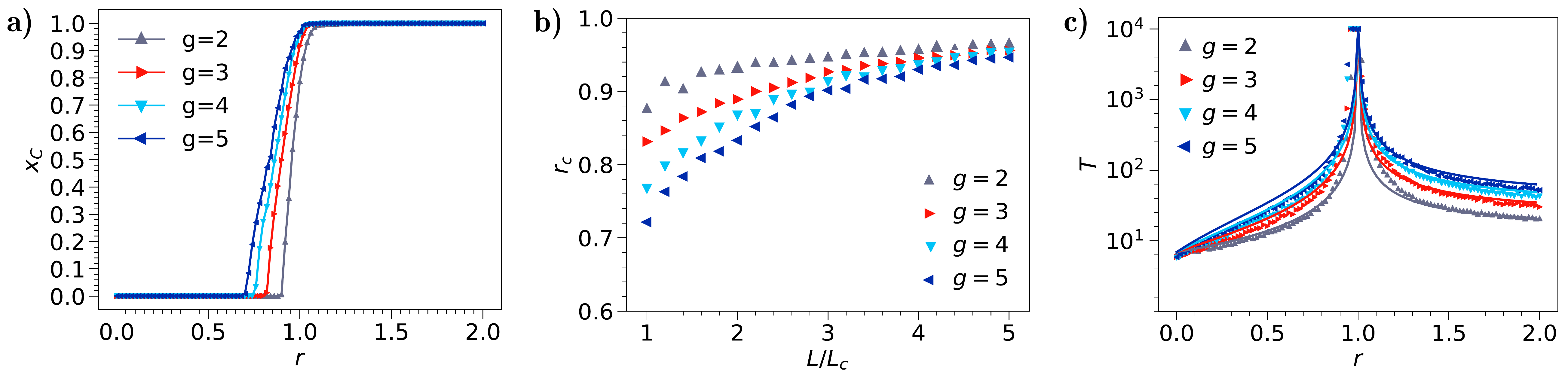}
\caption{Public Goods Game with higher-order interactions in uniform random hypergraphs. Numerical simulation of the Hypergraph Implementation of the game on uniform random hypergraphs of $N=1000$ players and different orders $g$. (a) Fraction of cooperators, $x_C$, as a function of the synergy factor, $r$, for hypergraphs with hyperdegree $\langle k\rangle=k_c$, or total number of hyperlinks $L=L_c$, where $k_c$ and $L_c$ stand for the critical hyperdegree or number of hyperlinks guaranteeing a connected hypergraph. (b) Critical value of the synergy factor, $r_c$, as a function of the ratio between the number of hyperlinks $L$ and the critical value $L_c$ in hypergraphs of different density. (c) Relaxation times as a function of the synergy factor, $r$, for hypergraphs with hyperdegree $\langle k \rangle=5 k_c$. In all plots, triangles correspond to numerical simulations, while the solid lines are the results of our theoretical predictions.}
\label{fig:fig2}
\end{figure}

\subsection*{Uniform Hypergraphs}
To get some insights into the dynamics of the system in a simple configuration, we first studied the PGG on uniform random hypergraphs (URH)
with hyperlinks of order equal to $g=$2, 3, 4 and 5 (see Methods for details on how to generate URH). Numerical simulations have been carried out for hypergraphs with $N=1000$ nodes (players), and the game has been iterated for $T=10^4$ time steps. The results obtained are reported in Fig. \ref{fig:fig2}. Panel (a) shows the final fraction of cooperators as a function of the reduced synergy factor $r$. In each case, the simulations refer to hypergraphs with $L=L_c$ hyperlinks, where $L_c$ accounts for the minimal number of hyperlinks that guarantees the connectedness of the hypergraph. As it can be seen in the figure, there is a value of $r$ beyond which cooperation emerges. We define this critical value of the reduced synergy factor, $r_c$, which depends on $g$, as the lowest value of $r$ for which the fraction of cooperators is nonzero. 

The results show that $r_c$ decreases when the order $g$ of the
hyperlinks of the hypergraph increases. This is equivalent to say that
$r_c$ decreases when the same number of $N=1000$ individuals play in
larger groups. We believe that this observation is important, since
determining how $r$ varies with the size of the group, allows us to get
more realistic insights. Admittedly, the well-mixed limit of
population-size groups is rarely applicable in reality, thus, the
study of the impact of having large groups inside large populations,
as allowed by our higher-order framework, is key. The panel (b) of
Fig. \ref{fig:fig2} displays how the value of $r_c$ depends on the number
of hyperlinks $L$ in the hypergraphs. For each value of $g$, we
observe an increase of $r_c$ with $L$, and a tendency, for large
hypergraph densities, to the value $r_c=1$, which corresponds to the
well-mixed replicator approximation
\cite{taylor_p_mb78}.
The replicator equation approximation relies
  on the indistinguishability of the nodes, and as such, it is 
  exact when the hypergraph is fully connected, i.e.~contains all the possible hyperlinks. However, we show that the
  approximation is good also for sparse hypergraphs, with a number of
  hyperlinks of the order of the critical value for ensuring a giant
  component. Therefore it is natural that the higher the value of $L$,
  the closer $r_c$ is to $1$. The same argument can be used to
  explain the results in (a). The ratio $L_c/C^{N}_{g}$, which
  represents the fraction between the critical number of hyperlinks $L_c$ and the
  total possible number of hyperlinks, given by the binomial coefficient $C^N_g$, decreases with $g$. This implies
  that, if two hypergraphs have $L=L_c$, but different values of $g$,
  the one with lower $g$ will be denser, and thus will exhibit a
  critical point closer to the analytic prediction. Therefore,
  we can say that at fixed reduced synergy
  factor $r$, large groups are better to foster cooperation in sparse
  hypergraphs, as the number of hyperlinks required for connecting all
  the players represents a smaller fraction of the total number of
  hyperlinks.
Finally, the value of $r$ also influences how long it takes for the
system to converge to the stationary solution. This is illustrated in
panel (c), where we report the measured relaxation time $T$
from an initial configuration with $x_D=x_C=0.5$, in a hypergraph with
$L= 5 L_c$. These results are obtained by running the simulations up
to a maximum of $10^4$ steps. Furthermore, for the replicator
approximation, the value of $T$ can be analytically computed as
$T=\frac{\ln(N-1)}{|Q|}$, with $Q=(1-r)/\Delta$ (see Equations \eqref{dyn} and \eqref{time} for the details of the calculation). As it can be seen in the
figure, the agreement between the theoretical predictions and the
numerical results is not only qualitatively but also quantitatively
very good.
The absorbing state, either full cooperation or full defection,
  emerges when the system is at equilibrium, a condition that can only be
  reached if enough iterations have occurred. On the other hand,
  real-world social
  interactions that can be modelled as games usually take place
  over a limited time interval $\tau$. Hence, 
the relation between the relaxation time $T$, which depends on the synergy factor $r$, and $\tau$ is crucial to  
determine if the system does or does not reach the equilibrium,
and consequently, if the replicator dynamics can or cannot predict
the numerically computed fraction of cooperators.
All these results indicate that the dynamics of the
PGG on uniform random hypergraphs corresponds to the replicator
dynamics in the well-mixed limit. In order to test
  the robustness of these findings with respect to the implementation
  selection, we have also carried out numerical simulations in the BI
  implementation (see Figure \ref{fig:di}).\\
         
\subsection*{Hyperdegree-Heterogeneous Hypergraphs}

The previous section addressed the simplest scenario 
in which the individuals of a population are assumed to be 
indistinguishable (URH). However, such an assumption can be too
oversimplified to describe real situations as it is well
known that social systems are heterogeneous. Think of your friends at college. It is likely that a minority of those are involved in considerably more activities, and therefore social circles, than the rest. Such heterogeneity is typically characterised by a
non-exponential degree distribution, allowing the presence of hubs, or
highly connected individuals \cite{latora2017complex}. 
Hence, we consider here two families of hyperdegree-heterogeneous hypergraphs that
we name power random hypergraphs (PRH) and scale-free random hypergraphs (SRH).
The algorithms we have used to generate these 
hypergraphs are reported in the Methods section, and their properties have been studied in Figure \ref{fig:urhben} and Figure \ref{fig:rhben}. Scale-free hypergraphs are characterised by a power-law distribution, and
represent the most hyperdegree-heterogeneous family of hypergraph considered
here. For this reason, these hypergraphs display a hierarchy between the nodes, as a few of them are involved in most of the hyperlinks and thus have a dominant position in the dynamics of the system. In contrast, power random hypergraphs stay in between uniform
and scale-free hypergraphs, as their hyperdegree distribution combines
exponential and non-exponential functions. 

To study the emergence of cooperators in
  hyperdegree-heterogeneous hypergraphs, we have run $T=10^4$ time
  steps of the game on
  ensembles of hypergraphs with $N=1000$ nodes and 
  orders $g=2,3,4,5$, respectively sampled from PRH and SRH.
In order to compare the
  simulations with those reported in Fig \ref{fig:fig2}a, we have fixed
  the total number of hyperlinks to $L=L_c$. When, for high
  heterogeneity, some of the nodes (a minimal fraction of the total) do not belong to the main component,
  we have neglected their contribution to the fraction of
  cooperators.

\begin{figure}[h]
\includegraphics[width=\textwidth]{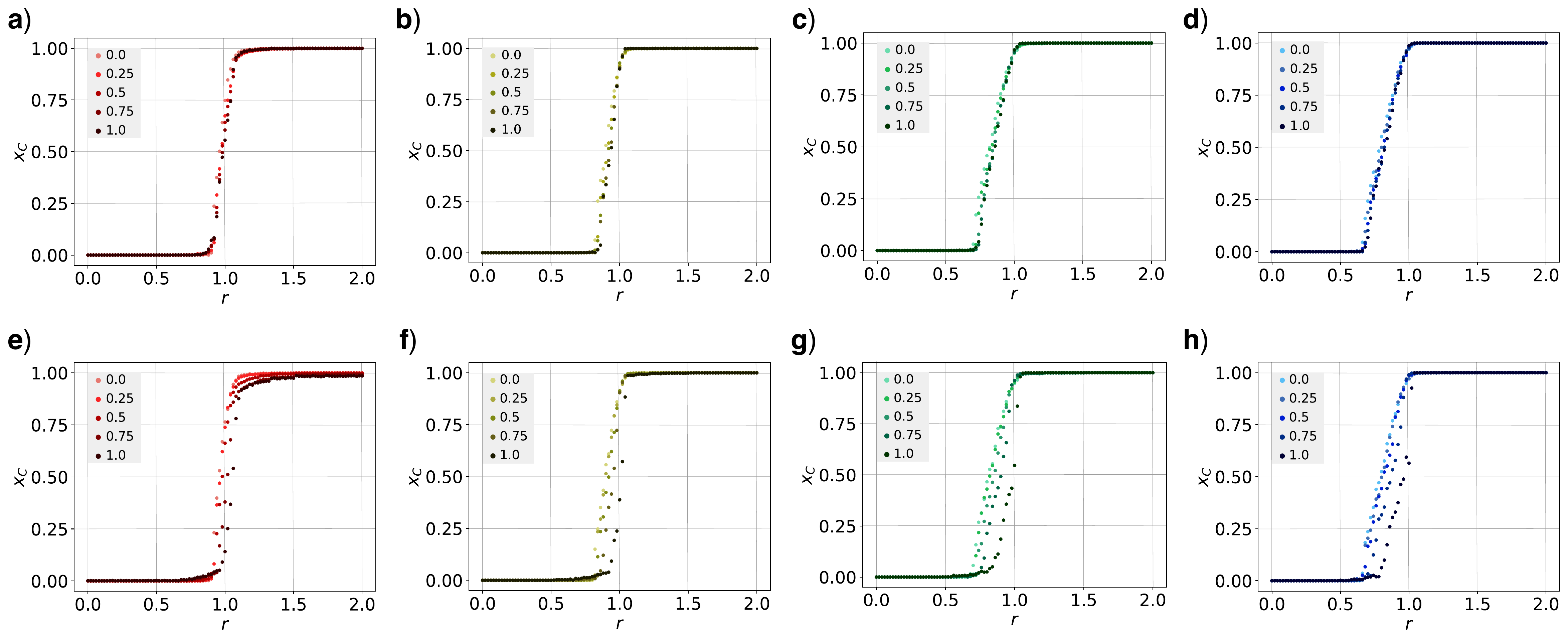}
\caption{Public Goods Game with higher-order interactions in hyperdegree-heterogeneous random hypergraphs. Numerical simulation for the fraction of cooperators $x_C$ as a function of the reduced synergy factor $r$ on hyperdegree-heterogeneous random hypergraphs of $N=1000$ players and different orders $g$. Top and bottom panels refer respectively to PRHs and scale-free random hypergraphs Random Hypergraphs (SRH), while the four different columns (a)-(e), (b)-(f), (c)-(g) and (d)-(h) correspond to the values of $g=2,3,4,5$. Legends in the PRH and SRH plots denote the value of $\mu$ characterising the hyperdegree distribution, where larger values $\mu$ imply higher heterogeneity.}
\label{fig:fig3}
\end{figure}

The results reported in Fig. \ref{fig:fig3} show an
  important difference between PRH and SRH. In the case of PRH
   (top panels) 
  the position of the transition does not depend on the heterogeneity
  of the node hyperdegree distribution, tuned by parameter $\mu$,
  (see Methods for the precise definition of $\mu$), 
  and the critical point is the same as that obtained in URH.  
  Conversely, the simulation of the game on SRH (bottom panels) shows 
  that, the larger the heterogeneity
  in the hyperdegree distribution (larger values of $\mu$),
  the more the solution deviates from that of URH, and the 
  closer the critical point gets to $r=1$. This indicates that 
  hierarchically structured systems inhibit cooperation in the PGG with higher-order interactions at variance with numerical simulations obtained on traditional networks under the same evolutionary dynamics.

In order to be able to explain these results 
we need to consider a refinement of the replicator approximation
that takes into account the possible presence of correlations between the
  hyperdegrees of nodes belonging to the same hyperlink.
Let $\mathcal{K}$ be the set of all possible
  hyperdegrees a node can have, and let $k \in \mathcal{K}$ be
  the hyperdegree of a randomly chosen node. We now denote as  
  $p({\bf k''}|k)$ the conditional probability that the node of
  hyperdegree $k$ is part of a hyperlink 
  where the remaining $g-1$ nodes
  have hyperdegrees ${\bf k''}=\{ k_1, k_2,\ldots, k_{g-1} \}$,
  where ${\bf k''} \in \mathcal{K}^{g-1}$ is a vector whose
 $g-1$ components are elements of $\mathcal{K}$. We have been able 
  to show that the system will fulfil the replicator approximation as
  long as the conditional probability $p({\bf k''}|k)$ does not depend on
  $k$ 
  (see Equation \eqref{hetmf} and the section below it for the detailed
  analysis). This is true for the case of the PRH.
  Conversely, in the case of the SRH, increasing heterogeneity
  while maintaining the total number of hyperlinks in the hypergraph
  requires reducing the number of effective nodes. This
  induces non-trivial correlations in the model 
  between the hyperdegrees of nodes
  belonging to the same hyperlink, and has a similar
  effect of driving the system closer to the $r_c=1$
  threshold, as that we have observed when we increase
  the hyperlink density in the uniform case (URH). Intuitively this can be explained by the notion of locality. When the density is low, or when no large hubs are present in the system, there is a non-negligible probability that cooperator bubbles emerge below the critical threshold, because there may be regions of the hypergraph that are semi-isolated, and therefore protected from defectors, even if they belong to the same component. However, either increasing the density or introducing hubs will reduce the probability of finding these isolated groups of nodes, and therefore will inhibit the formation of cooperator bubbles below $r=1$.

\subsection*{Order-Heterogeneous Hypergraphs}
Heterogeneity can also arise in the order of the hyperlinks. Indeed, the proposed HI of the PGG allows studying the more general, realistic
and interesting case of hypergraphs where not all the hyperlinks have
the same order. Important examples of such systems include teams of
different sizes working for a common goal or one-to-many communication
via apps like WhatsApp, where users can create and belong to several
groups of different sizes. In what follows, we consider order-heterogeneous
random hypergraphs with an assigned distribution of hyperlinks. Such
hypergraphs are characterised by their total number of hyperlinks $L$
and by a probability vector ${\bf p}= \{ p^g \}_{g=g^-}^{g^+}$, whose
entry $p^{g} = k^{g}/k$ specifies how likely it is, on average, that
the hyperdegree $k$ of the node contains $k^{g}$ groups of order
$g$. ${\bf p}$ is normalised such that $\sum_{g=g^-}^{g^+} p^g =
1$. Considering groups of different orders in the same hypergraph
allows us to focus on another important aspect of the PGG on higher-order
structures, namely, the possible dependence of the rescaled synergy
factor $r$ on the order of the group. This is important for practical
purposes, given the increasing interest in understanding how the size
of a group impacts its performance. As it has been shown recently
\cite{dashun2019}, large and small teams play different roles in
science and technology ecosystems. Thus, it is natural to assume that
the synergy factor of a group depends on its size. This is
particularly true in scientific publications, where it has been shown
that the larger the group, the more citations a produced publication is
likely to attract \cite{wuchty2007,klug2016}. Therefore, as a 
general form for such a dependence we assume that the synergy factor $R$ is an
increasing power-law function of $g$, namely:
\begin{equation}
R(g) =\alpha g^{\beta}
\label{eq_alpha}
\end{equation}
with parameter $\alpha>0$ and exponent $\beta \ge 0$. The value of the
exponent allows to tune the benefit that the players are able to
produce when working as a group. In particular, adopting a superlinear
scaling $\beta > 1$, means considering a synergistic effect of a group
that goes beyond the sum of the individual contributions
\cite{bettencourt07,bettencourt13}. Notice, however, that
the assumed dependence in Eq.\ (\ref{eq_alpha}) is only a first
approximation as it neglects saturation effects or
even possible disadvantages due to difficulties in coordinating large
groups, which, as we will see later on, appear in
real systems. 
Under this assumption, the average payoff difference between
cooperation and defection can be written as:
\begin{equation}
\pi_D - \pi_C = \sum^{g_+}_{g=g_-} p^{g} (1-\alpha g^{\beta -1})
\end{equation}
where $g_-$ and $g_+$ are again the minimal and maximal orders of hyperlinks, respectively. The relaxation time is again given by $T={\ln(N-1)}/{|Q|}$, where $Q=(\pi_D - \pi_C)/ \Delta$ (see the Appendix for the definition of $\Delta$ in the general case and for explicit calculations). It is then possible to derive the critical value of the parameter $\alpha$ as a function of the exponent $\beta$ as: 
\begin{equation}
\alpha_c (\beta)=\frac{1}{\sum^{g_+}_{g=g_-} p^g g^{\beta -1} }=\frac{1}{\mathcal{K}_\beta},
\end{equation}
where, for simplicity, we have defined $\mathcal{K}_\beta \equiv \sum^{g_+}_{g=g_-} p^g g^{\beta -1}$.
We remark here that $\alpha=\alpha_c$ for a fixed value of 
  $\beta$ is the critical point separating the
  defection and cooperation phases. This means that when $\alpha < \alpha_c$
  the system will converge to full defection, while for $\alpha
  > \alpha_c$ it will converge to full cooperation.

\begin{figure}[h]
\includegraphics[width=\textwidth]{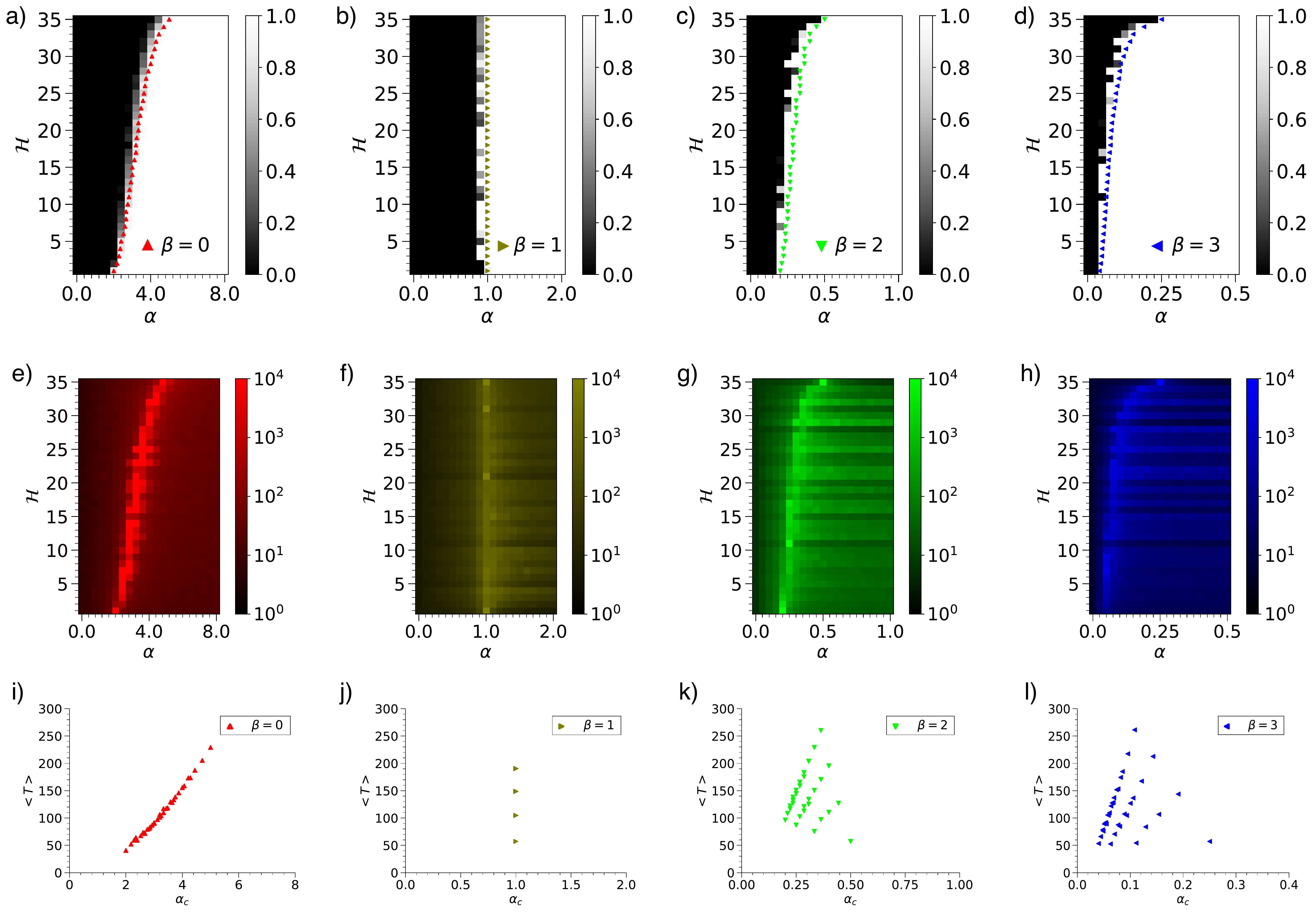}
\caption{Public Goods Game with higher-order interactions in order-heterogeneous random hypergraphs. We assume that the synergy factor grows according to Eq.\ (\ref{eq_alpha}) and consider the set of hypergraphs $\mathcal{H}_i$ that contain hyperlinks of orders $g=\{2,3,4,5\}$ with probabilities $p_g$ taking values in the set $\{0,0.25,0.5,0.75,1\}$ $-$there are 35 possible such hypergraphs. (a-d) Fraction of cooperators as a function of $\alpha$ for each of the 35 hypergraphs $\mathcal{H}_i$ and several values of $\beta$. The hypergraphs are ordered according to their value of $\mathcal{K}_\beta$. Simulations have been carried out up to $T=10^4$ time steps for hypergraphs with $\langle k\rangle=2 k_c$, and triangles correspond to the theoretical predictions in the replicator approximation (see the Appendix for details). (e-h) Relaxation time as a function of $\alpha$ for the set of hypergraphs $\mathcal{H}_i$. Now hypergraphs have $\langle k\rangle=5 k_c$. (i-l) Predictions for the critical value $\alpha_c$ as a function of the average relaxation time, calculated for each hypergraph in $\mathcal{H}_i$ by averaging over the intervals of $\alpha$ [0,8], [0,2], [0,1] and [0,0.5] for $\beta=0,1,2,3$ respectively.}
\label{fig:fig4}
\end{figure}

To explore how the dynamics evolves in order-heterogeneous random hypergraphs, we have performed numerical simulations of the PGG considering four different values of $g=2,3,4$ and $5$ and allowing the values of $p^g$ to take only multiples of $0.25$. This leads to 35 possible hypergraphs, one for each of all conceivable convex sums of $\{p^2,p^3,p^4,p^5\}$ with the previous constraints.
This means that the hypergraphs we consider are composed
  by hyperlinks of different orders, where each order $g$ takes $L p^g$
   hyperlinks out of the total number $L$. For instance, on a hypergraph with
  $L=100$ and order probabilities $(0,0.25,0.25,0.5)$, on average we
  would expect $25$ hyperlinks of order $g=3$, another $25$ of order $g=4$
  and the remaining $50$ of order $g=5$.
  Results are reported in Fig. \ref{fig:fig4} for four different values of the power exponent $\beta$, namely, $\beta=0,1,2,3$, shown with different colours. Notice that the case $\beta=1$ corresponds to the underlying linear assumption of the standard PGG: in this case, $\alpha$ plays the role of the reduced synergy factor $r$. Panels a) through d) plot the colour-coded fraction of cooperators as a function of the parameter $\alpha$ in the definition of the synergy factor. The hypergraphs $\mathcal{H}_i$ have $\langle k\rangle=2 k_c$ and are displayed according to their value of $\mathcal{K}_\beta$, i.e.,
the value of the critical point $\alpha_c(\beta)$. As for the case of uniform random hypergraphs, we find that although the critical point is slightly overestimated for low densities by the analytical approximation, there is still a good agreement between the theoretical predictions of the well-mixed replicator approximation and the numerical simulations. We next explore the behaviour of the relaxation time. Panels e) through h) show results obtained for order-heterogeneous hypergraphs with $\langle k\rangle=5 k_c$. As it was done for the homogeneous scenario, we follow the dynamics of the system up to a maximum of $T=10^4$ time steps. The plots show that the relaxation times depend on $\alpha$ for all values of $\beta\neq 1$, albeit rather differently with respect to the dependence of the critical value $\alpha_c$ for $\beta < 1$ and $\beta > 1$. In order to further explore this relationship, we analysed how the average relaxation time varies as a function of the critical point $\alpha_c$. Results shown in panels i) to l) reveal that the dependence is always linear. However, when the synergy factor increases super linearly, there appear different curves, each one corresponding to a distinct family of hypergraphs and characterised by a different linear relation between the average relaxation time and the critical value. This behaviour introduces an additional degree of freedom that can turn very useful, since the degeneracy that is observed for $\beta \le 1$ is broken for $\beta >1$, and therefore one can independently set a critical point and a relaxation time by opportunely choosing the corresponding hypergraph.
We remind the reader that cases with $\beta > 1$ are those in
  which the synergy factor $r(g)$ has a superlinear dependence on the
  order $g$.   
Those values of $\beta$ are \textit{a
    priori} the most interesting ones to study, and the ones more
  likely to be found in real situations. Therefore, our results about
  the relaxation are particularly relevant, because in this case
  one can potentially turn an
  unstable system into a stable one, and the opposite, by changing
  the order of the hyperlinks, while still respecting the value of the
  critical point.

\subsection*{Synergy factor of real games}
From the previous results, a natural question arises: is it possible
to determine the value of the synergy factor for a real PGG for each
of the possible group sizes? A plausible answer to this question can
be obtained under the assumption that the very same structure of the
hypergraph is the result of an evolutionary process in which nodes
select the groups they belong to. We hypothesise that each individual
tries to optimise the ideal number of groups of each order, based on
the perceived dependence of the synergy factor on the group size. In
this way, each real-world hypergraph would be the optimal structure
that supports the game it hosts. We could then extract the functional
form $R(g)$ directly from the hyperdegree distribution of the
hypergraph. More precisely, the goal would be to use the information
in the vector ${\bf p}$ of the hypergraphs on which the PGG occurs to
determine the functional form, $R(g)$, of the synergy factor by
imposing two conditions. The first condition comes from the assumption
that the unknown reduced synergy factor $r(g)$ is proportional to
$p_g$. This originates in the intuition that the distribution of the
hyperdegree of a generic player should be aligned with the potential
benefit that each player expects to obtain for each higher-order
interaction. The second condition imposes that the average payoff of
cooperators is equal to the average payoff of defectors. This implies
that the system is at equilibrium and guarantees the coexistence of
cooperators and defectors. Thus, given that these two conditions are
satisfied, it is possible to extract the curves of $r(g)$ and $R(g)$
from empirical data on higher-order interactions.

\begin{figure}[h]
\includegraphics[width=\textwidth]{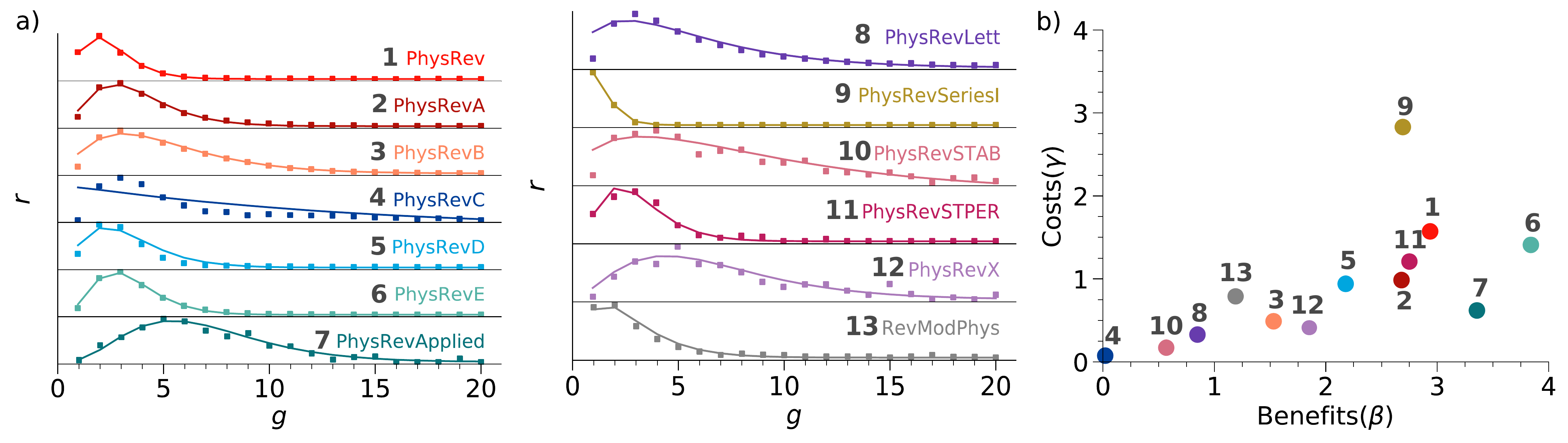}
\caption{Synergy factors of scientific collaborations. (a) Empirical synergy factors extracted from the structure of hypergraphs describing co-authorships from publications in the American Physical Society journals. Solid lines are fit of the empirical dots according to Eq. \eqref{fit}. Different symbols and colours refer to different journals, i.e. to different scientific communities. (b) Journal (labelled from 1 to 13) positions in the costs-benefits diagram. The synergy factors are factorized as a function of two competing terms, one modeling the benefits of cooperation, which is dominant for small group size, and another one accounting for costs associated to an excessively large number of co-authors, which describes for the exponential decays observed in the first two panels.}
\label{fig:fig5}
\end{figure}

In order to show how the above-mentioned procedure works in
  practice in real cases, we have studied collaboration in science and
  technology. We believe that this could constitute an example in which the benefit of a group depends on its size and at the same time, all group members do not contribute the same to the collective task, which essentially leads to a PGG dynamics. Although there is not a single way of classifying in a binary manner (either cooperator or defector) the authors of a scientific publication, one can think of two type of players mimicking cooperators and defectors. A cooperator can be considered as anyone that has contributed at least a ``fair" amount of work. The reverse applies to defectors, which can be considered those that put less effort in producing a teamwork than the average or the ``fair" amount of work. Note that whatever the effort of the team members is, they all receive the same benefit, for instance, in terms of citations (the citation is to the paper, not to the individual). Thus, given that there are cooperators and defectors, what is the optimal collaboration (group) size? And that of the synergy factor?
  
  In particular, we have considered a large data set of
  all the scientific articles published in the last century in
  thirteen journals of the American Physical Society (APS). For each
  journal, we have constructed a hypergraph whose nodes and hyperlinks
  represent respectively scientists and co-authored publications
  (see Table 1 for further details).  The reduced
  synergy factors have then been obtained from information on the
  number of authors in each publication (see Methods). From the plots
  of $r(g)$ vs $g$ reported in Fig. \ref{fig:fig5}a we notice the
  existence of a maximum value of $r$ at intermediate group orders $g$. 
  This indicates that there is an optimal trade-off between
  the positive and negative effects of increasing the group size. 
  The optimal value of $g$ depends on the specific scientific
  community, as it varies from journal to journal. In the case of PhysRevLett the maximum of $r(g)$ 
is located at $g=3$.
  Different journals are associated with other optimal
    collaboration sizes. For instance, for PhysRevApplied $r(g)$
    is maximum at $g=5$, indicating that larger collaborations are 
    more beneficial in applied topics, such as
    device physics, electronics and industrial physics. For almost all
    journals, the synergy factor is low for $g=1$, showcasing the
    difficulty of publishing alone in physics, a research area where
     teamwork has been becoming increasingly
     important in the last decades~\cite{battiston19}.
     Interestingly, a paradigmatic case
    is the one of PhysRevSeriesI, the very first journal 
    published by the APS in the early 1900s, for which a sharp peak is located at $g=1$,
    showing how most publications where produced by single
    scientists, in contrast with current trends.  In order to shed
  light on this result we have factorised the synergy factor as the
  product of an increasing function of $g$ times a decreasing function
  of $g$, and we have performed a numerical fit to extract the
  benefit exponent $\beta$ and the so-called cost parameter $\gamma$
  (see Methods).  This enables us to interpret the synergy factor as a
  combination of two contrary effects of the higher-order interactions
  in this particular dataset.

Fig. \ref{fig:fig5}b reports the values of $\beta$ and $\gamma$ 
  obtained
  for each journal of the APS, and it allows us to classify the
  different scientific communities in terms of 
    benefits and costs of higher-order interactions. These results
  provide a game-theoretic interpretation of the APS dataset.
  Specifically, in the context of this
  bibliographic dataset, hidden benefits and costs that conform to the
  synergy factor can be associated with several aspects of the task of
  producing a publication.  Benefits (an increase of the synergy factor
  with increasing $g$) would correspond to the potential reinforcement
  of the amount and quality of the ideas and the potential increase in
  the outreach of the work with the number of co-authors involved. On
  the contrary, the costs (decrease with increasing $g$) would be the
  additional organisational effort in the process of arriving at a
  consensus and carrying out the tasks for publishing a paper.
  Experimental communities, such as that of nuclear physicists publishing
  in PhysRevC, tend to have low costs. These
  ideas are aligned with recent studies about the creation and
  production of research ideas \cite{milojevic} and the
  role, group dynamics and success of teams
  \cite{dashun2019,wuchty2007,klug2016}. Our formalism allows for a
  quantitative analysis of these phenomena and could be used in future
  applications to design ways to foster higher-order cooperation.

\section*{Discussion}
Summing up, we have introduced higher-order interactions in evolutionary games to study cooperation in groups. Since higher-order interactions allow for a single link to connect more than just two individuals, they are naturally suitable to define groups in networks. In doing so, higher-order interactions thus do away with the arbitrary definitions of groups in classical networks, and they provide a formal theoretical foundation to study cooperation in networked groups. We have shown that the public goods game on a hypergraph is effectively a null model that agrees exactly with the replicator dynamics in the well-mixed limit as long as no hyperdegree-hyperdegree correlations exist. As such, it can be used in future research towards upgrades that add additional layers of reality in models of human cooperation, either by means of strategic complexity \cite{perc_pr17}, or by means of more complex interaction networks \cite{wang_z_epjb15}. 

Towards the latter effect, we have also studied the public goods game
on hyperdegree-heterogeneous and on order-heterogeneous hypergraphs,
where we study the effects of the presence of highly
  connected individuals and of hyperlinks of different orders  
  respectively. Due to the exact definition of a group in the
proposed framework, we have been able to systematically and
consistently consider synergy factors that are dependent on group
size. Indeed, the framework allows us to unveil the effects of group size 
on cooperation in its most general form. As an example, we have
considered synergy factors that are given by different powers of the
group size, showing a critical scaling in the transition from
defection to cooperation. In this case too, we have observed a
substantial agreement between the simulations and the analytical
predictions of the model. Interestingly, we found that hierarchically structured hypergraphs could hinder cooperation in a structured population. Our framework enables analysis of real
systems, as we have shown for the APS publications dataset, providing
insights regarding the positive and negative effects associated to
higher-order interactions and the nature of group dynamics. However, even if our framework includes diverse forms of higher-order interactions, we recognize that a current limitation of this representation of human interactions is given by the constraints imposed by the available data. Admittedly, the identification of interactions in social networks beyond the traditional pairwise relationships constitutes nowadays an important challenge. Interestingly enough, this also represents an opportunity from an experimental point of view. It is also worth mentioning that the application of our results to scientific publications is based on the hypothesis that the interaction structure is the outcome of an optimization process, where the average distribution of groups that each node is part of coincides with the synergy factor, such that the system is in a stationary state of the dynamics of the PGG. This hypothesis, which constitutes a limitation of our method to extract synergy factors from real data, could potentially be either validated or refuted by models considering the dynamics within the topology of interactions on top of the PGG. Moreover, the PGG imposes all cooperators to contribute with the same amount, making this contribution a boolean variable in practice. And therefore, an additional limitation arises when adapting real systems to the rigid formalization of the PGG, as the role of cooperators and defectors cannot be unequivocally defined when the contributions are not only 0 or 1.

It is also worth mentioning that in his essay titled Innate Social
Aptitudes of Man, W.~D. Hamilton wrote, ``There may be reasons to be
glad that human life is a many-person game and not just a disjointed
collection of two-person games''. He was referring to the fact that
social enforcement works better in groups with more than two members,
which can offer at least a partial cure for the problems with
reciprocation in larger groups \cite{sigmund_tee07}. We note that the theoretical framework of
  higher-order interactions also invites to re-examine other fundamental mechanisms that may promote cooperation, such as image
scoring \cite{nowak_n98, milinski_prslb01, nax_srep15b}, rewarding
\cite{rand_s09}, and punishment \cite{andreoni_aer03, gaechter_s08,
  boyd_s10, jordan_n16}.

Given the fundamental differences between pairwise and higher-order
interactions, it would also be of interest to revisit the role of
specific network properties and their role in the evolution of
cooperation.  In this regard, the role of community structure
\cite{fotouhi_rsif19}, as well as two or more network layers
\cite{wang_z_epl12, gomez-gardenes_srep12, gomez-gardenes_pre12,
  wang_z_srep13, wang_pre14a, battiston_njp17, fu2017leveraging,
  fotouhi2018conjoining}, promise to be fruitful ground for future
explorations on how interaction structure impacts
cooperation. Overall, we believe that the introduction of higher-order interactions to evolutionary games has the potential to improve our understanding of the evolution of cooperation and other social processes in networks.

\section*{Methods}

\paragraph*{Uniform random hypergraphs (URH):} 
We detail here the procedure we have adopted to sample
  $g$-uniform hypergraphs, i.e. hypergraphs with all hyperlinks of the
  same order $g$. A URH of order $g$ can be constructed by assigning a
uniform probability $p$ to each $g$-tuple of $\mathcal{N}$. For each
of them, a random number in the $[0,1)$ interval is generated, and if
  this number is lower than $p$, the hyperlink containing the
  $g$-tuple is created. However, this method scales badly with $g$
  since the number of $g$-tuples to be considered is equal to the
  binomial coefficient $C^{N}_{g} = {{N}\choose{g}} $, which grows
  fast as a function of $g$.  A more efficient procedure is to fix the
  total number of hyperlinks, $L$, and generate a random integer in
  the $[1,C^{N}_{g}]$ interval.
One has to provide an 
  ordering for the set of all possible hyperlinks, so that each of the
  random integers corresponds to a hyperlink. The
  hyperlinks selected through this process are then added to the hypergraph.
  The hyperlink ordering is based on the
    following combinatorial identity
\begin{equation*}
C^{N}_{g}=\sum^{N-(g-1)}_{i=1} C^{N-i}_{g-1}
\end{equation*}
that allows us to partition set $\mathcal{L}$
of all the possible hyperlinks of a $g$-uniform
hypergraph in terms of disjoint hypergraphs,
each one of them containing the hyperlinks that form the corresponding g-star hypergraph \cite{eliad88}.
This holds true in general, which enables us to apply the same
argument recursively, such that we can order all the possible hyperlinks
univocally, and even more, the probability for having a specific node
in a hyperlink is equal for all the nodes. These properties arise from the combinatorial
probabilities $d_i = \frac{C^{N-i}_{g-1}}{C^{N}_{g}}$ for $i=1,
\ldots N -(g-1)$ i.e.
the normalised weights of each of the terms
in the summation. We have empirically found a
distribution that can be used as an approximation to
$c_i = \sum_{j=1}^i d_j$, the cumulative distribution of $d_i$, namely
given by $1-(1-x)^g$, where
$x=i/(N-(g-1))$. See Figure \ref{fig:urhben} for a numerical proof 
of the convergence between both expressions.

  For the purpose of studying the stationary condition of a game, we are
  interested in having a connected hypergraph. The critical thresholds
  for the number of hyperlinks, $L_c$, and the hyperdegree, $k_c$,
  are equal to $L_c=\frac{N}{g} \ln{N}$ and $k_c = \ln{N}$.
  Hence, when $L$ is larger than $L_c$, 
  there is a high probability that the resultant hypergraph is
  connected.

\paragraph*{Power random hypergraphs (PRH):} We have
  seen that using the combinatorial probabilities $d_i$ allows us to
  create uniform random hypergraphs. Therefore, increasing the value
  of the exponent $g$ to $g'$ in $c_i$, such that $g'>g$, will
  increase the probability of sampling the hyperlinks belonging to the 
  g-star hypergraphs of low index nodes, and therefore introduce
  heterogeneities in the degree distribution. The control parameter
  that we use in the simulations in the manuscript is $\mu \in
  [0,1]$. In terms of $\mu$, one can obtain the power to use in the
  cumulative distribution $g'$ as $g'=(1+\mu)g$. In order to sample
  hyperlinks of order $g$ according to the new distribution, we
  transform the random number $r$ to a different random $r'$
  \begin{equation}
r'=c_{i-1}(0)+[r-c_{i-1}(\mu)]\frac{d_i(0)}{d_i(\mu)}
\end{equation} 
Here $i$ is the g-star to which the hyperlink would belong if it was sampled according to $r$. In this expression $d_i(\mu)$ and $c_i(\mu)$ account for the distributions using the value of $g'$ as a function of $\mu$. Accordingly, $d_i(0)$ and $c_i(0)$ are simply the distributions of the uniform case. See Figure \ref{fig:rhben} for the analysis of degree distribution emerging from the PRH.

\paragraph*{Scale-free random hypergraphs (SRH):} The standard indicator of heterogeneity in graphs is the power-law decay of the degree distribution. Here we employ the static scale-free algorithm \cite{jhun19} to generate such a profile. We use the same control parameter as in the PRH, $\mu \in [0,1]$, which in this case results in a power-law $p_k \sim k^{-\lambda}$ where the power $\lambda$ is $\lambda=1+1/\mu$. See Figure \ref{rhben} for the degree distribution of the hypergraphs generated with the SRH.

\paragraph*{Extracting synergy factors from real data:} 
We show here how the dependence of the reduced synergy factor $r(g)$
on group size $g$ can be derived for real systems, based on the
assumption that this information is encoded in the very same structure of
a hypergraph. In particular, we have considered a data set of
scientific publications and we have used it to investigate how
benefits change with the size of groups in scientific
collaborations. The data set consists of 577886 papers published in
the period from 1904 to 2015 in the collection of all the journals of
the American Physical Society (APS) \cite{apsdata}.
We have constructed the 13 hypergraphs corresponding to 
different journals, such as Physical Review, Physical Review
Letters, etc. of the APS. 
The nodes and hyperlinks of these hypergraphs represent scientists
and publications respectively. The order of a hyperlink
is equal to the number of authors of the
corresponding publication.  For each hypergraph, we have extracted
the number $L^g$ of hyperlinks of a given order $g$, which we used to compute the
average number $k^{g}=g L^{g} /N$ of hyperlinks of order $g$ a node is
involved in. The reduced synergy factor $r(g)$ can then be
extracted from the proportion $p^g = k^g/k$ of hyperlinks of order $g$
of a node, by assuming that $r(g) = z p^g$ and using the critical
point relation:
\begin{equation}
\sum^{g=g^+}_{g=g^-} p^g (1-r(g)) = 0
\end{equation}
to calculate the proportionality constant $z$.

\paragraph*{Cost-benefit factorisation of the synergy factor:}In scientific collaborations across all journals of the APS, an optimal team size is associated with a maximum in the synergy factor, suggesting that an excessively large number of co-authors might lead to disadvantages in cooperation. In order to account for these effects, we have modelled the synergy factor extracted from real-world collaboration data as the following function of $g$: 
\begin{equation}
f(g,\alpha,\beta,\gamma)=\alpha g^{\beta} e^{-\gamma (g-1)}.
\label{fit}
\end{equation}
ruled by the three parameters, $\alpha, \beta$ and $\gamma$. The first parameter, $\alpha$, introduced in Eq. \eqref{eq_alpha}, is determined by the critical point condition. The remaining two parameters account respectively for the benefits and costs of the higher-order interactions. Benefits are modelled as a power-law of the group size $g$ with an exponent $\beta$. Costs are described by an exponential decrease in the group size tuned by the cost parameter $\gamma$. Different functions of $g$ might also provide a satisfactory fit of the data. Here we have opted for this expression because it enables to factorize the group size dependence into two different contributions, benefits and costs, that can be interpreted in terms of behaviors of the players. The benefits grow as $g^{\beta}$, where $\beta$ captures the synergistic effect of group interactions. The term due to the cost associated to 
task organization in groups has its maximum at $g=1$,  and the exponential dependence has been adopted to avoid possible singularities of other functional forms at $g=1$. 
In conclusion, Eq. \eqref{fit} has a maximum at $g=\beta/\gamma$, which summarizes the result in a compressed expression. To extract the pair of parameters $(\beta,\gamma)$ for each journal, we have explored the parameter space and performed an optimisation in order to reproduce the empirical points correctly. For each considered pair $(\beta,\gamma)$, we have computed the normalised distance between the synergy factor inferred analytically and the one associated to the data (see Equation \eqref{last} for further details on the procedure). The pairs with the smallest distance are selected as the outcome of the optimisation process and are those reported in Fig. \ref{fig:fig5}c.

\noindent \textbf{Acknowledgements} \\
U.A.-R. acknowledges support from the Spanish Government through Maria de Maeztu excellence accreditation 2018-2022 (Ref. MDM-2017-0714) and
from the Basque Government through the Postdoctoral Program (Ref. POS-2017-1-0022). F.B. acknowledges partial support from the ERC Synergy Grant 810115 (DYNASNET). V. L. acknowledges support from the Leverhulme Trust Research Fellowship ``CREATE: the network components of creativity and success''. Y. M. acknowledges partial support from the Government of Arag\'on and FEDER funds, Spain through grant E36-20R to FENOL, by MINECO and FEDER funds (grant FIS2017-87519-P) and from Intesa Sanpaolo Innovation Center. M. P. was supported by the Slovenian Research Agency (Grant Nos. J1-2457, J1-9112, and P1-0403). G. F. A. acknowledges support from Intesa Sanpaolo Innovation Center. We thank M. Clarin from COSNET Lab for help and assistance with the figures. The funders had no role in study design, data collection and analysis, decision to publish, or preparation of the manuscript.

\noindent \\ \textbf{Author contributions} \\
U. A.-R, F. B. and V. L. conceived the study with contributions from G. F. A, M. P and Y. M. U. A.-R performed the calculations, U. A.-R, F. B., G. F. A, M. P, Y. M. and V. L. analysed the data and discussed the results. U. A.-R, F. B., G. F. A, M. P, Y. M. and V. L. wrote the paper.

\noindent \\ \textbf{Competing interests} \\
The authors declare no competing interests.

\noindent \\ \textbf{Data availability} \\
The APS dataset is provided by the APS at their website: https://journals.aps.org/datasets.

\noindent \\ \textbf{Code availability} \\
Custom code that supports the findings of this study is available from the corresponding author upon request.

\clearpage

\section*{Appendix}

\subsection*{Game Implementations}
\paragraph*{Graph Implementation}
In the original graph implementation, or GI, the core idea is to derive the higher-order structure from the network of players. In the GI every node constructs its own focal hyperlink, whose constituents are the node itself and its first neighbours. Therefore, for a system of $N$ players the GI imposes $N$ different groups, each of them of $g=k^2_i +1$ members. A micro-step in the game is played by following a sequence of steps: Firstly a node, and one of its neighbours, say $n_i$ and $n_j$, are randomly selected from $\mathcal{N}$. Secondly, the nodes play a round of the game for all the groups they belong to, accumulate their payoffs, $\pi_i$ and $\pi_j$, and normalize them with the number of games they have played. As there is a group for each player every node will play $k^2_i +1$ rounds. The third step is the update of the strategy, and for that purpose we employ the replicator update rule. According to this, $n_i$ will adopt the strategy of $n_j$ with probability $\frac{1}{\Delta}(\pi_{j}-\pi_{i})\theta (\pi_j -\pi_i)$, where $\Delta$ is the absolute value of the maximal payoff difference for all the possible strategies. It is noteworthy to mention that the replicator update rule is not the only update mechanism, but it is the one that we have selected because it provides an effective and noisy payoff oriented update rule by means of a simple analytic expression. 

\paragraph*{Bipartite Implementation}
The Bipartite Implementation (BI) lies at an intermediate point between the GI and the HI. On the one hand, and like in the HI and unlike in the GI, each round of BI occurs in a hyperlink of a hypergraph, which makes the implementation consistent with the structure of higher-order interactions present in the system. On the other hand, the update process does not benefit from the simultaneous interaction with multiple players, as each node updates its strategy by comparing its payoff with a randomly selected neighbour. The algorithm would be as follows: at each step of the game a node, one of its hyperlinks and a neighbour belonging to this hyperlink are randomly selected. Both nodes play a round of the game for each hyperlink they are part of. The original node updates its strategy by comparing its payoff with the one of its randomly selected neighbour.\\

\begin{figure}[h]
\captionsetup[subfigure]{labelformat=empty}
\centering
\subfloat[(a)]{\includegraphics[width=0.5\textwidth]{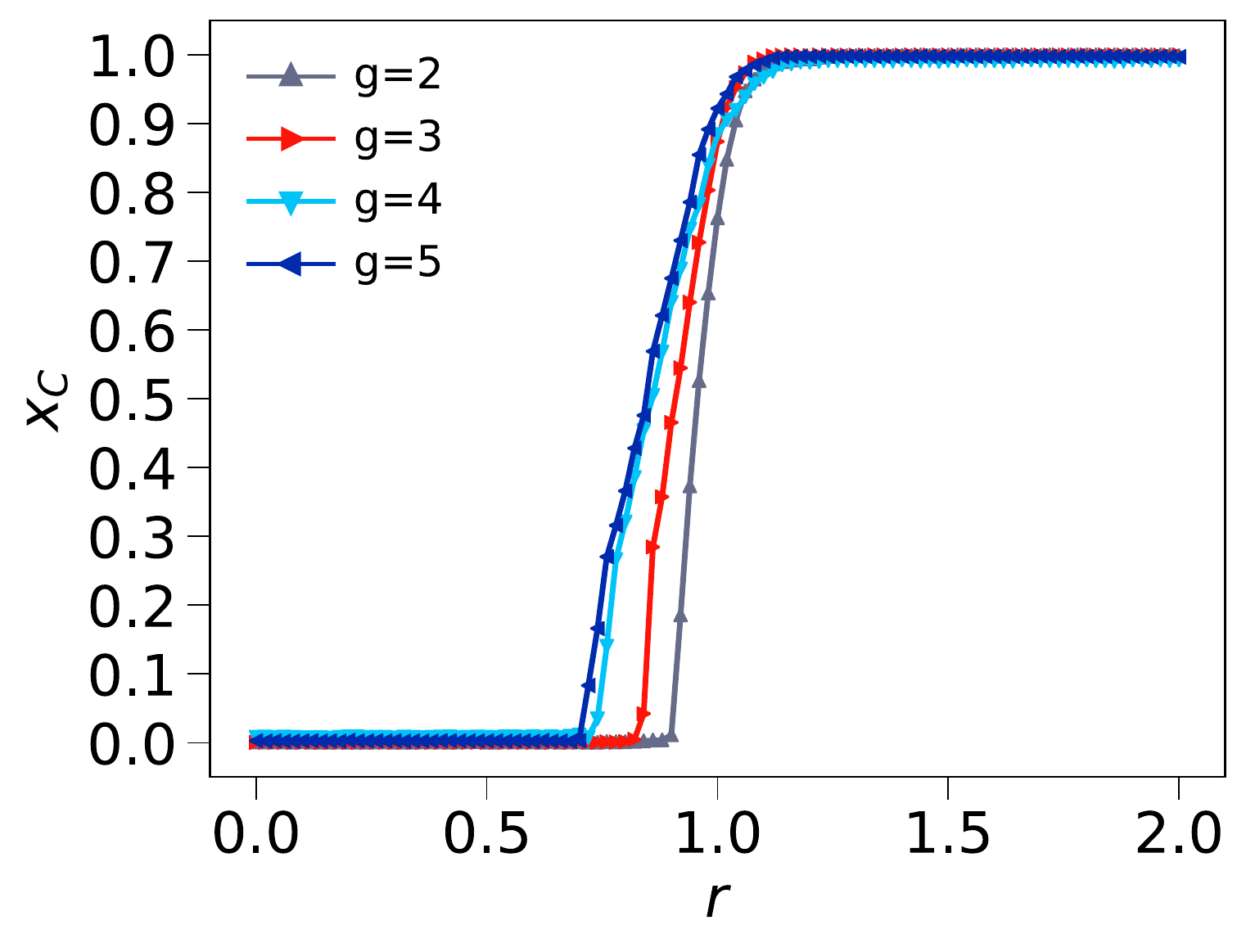}}
\subfloat[(b)]{\includegraphics[width=0.5\textwidth]{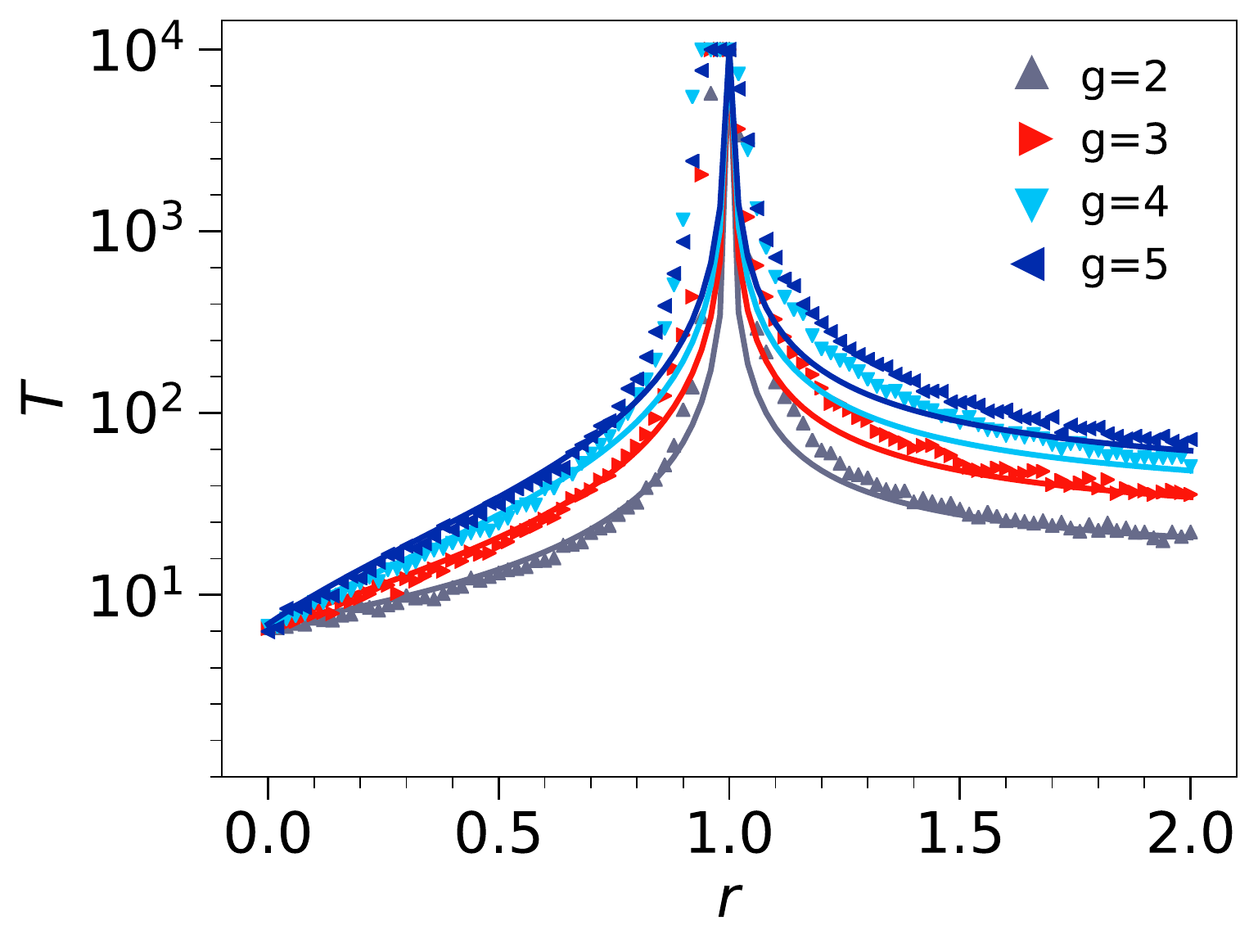}}
\caption{Public Goods Game in the Bipartite Implementation. (a) We depict the fraction of cooperators as a function of the synergy factor for a system with $N=1000$ nodes, and $L=L_C$ hyperlinks evolving during $T=10^4$ time steps. (b) We show the relaxation time of the system as a function of the synergy factor for a hypergraph with $N=1000$ nodes and $L=5 L_C$ hyperlinks.}
\label{fig:di}
\end{figure}

We have carried out a series of numerical simulations in an ensemble of hypergraphs with $1000$ nodes for different values of the group size $G$, see Figure \ref{fig:di}. We first observe that the critical point is the same as in the HI, meaning that the asymptotic emergence of the defectors or cooperators is the same in both cases. However we do observe a difference in the relaxation time: even if HI and BI are comparable, proving the robustness of the procedure, BI is slower, and it exceeds the limit established by the replicator approach.

\subsection*{Uniform Random Hypergraphs}
\subsubsection*{Creation Algorithm}
\begin{figure}[h]
\captionsetup[subfigure]{labelformat=empty}
\centering
\subfloat[(a)]{\includegraphics[width=0.5\textwidth]{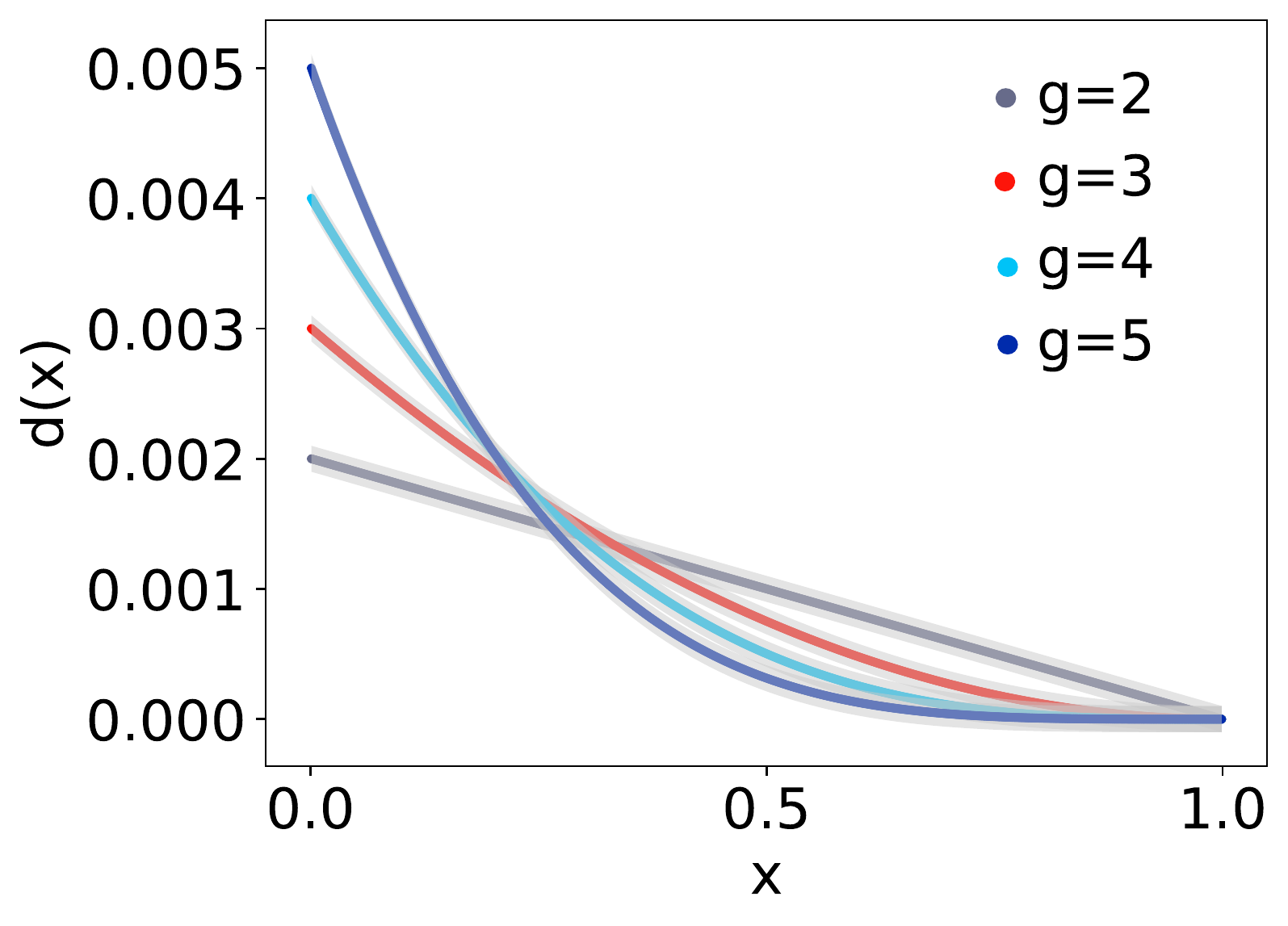}}
\subfloat[(b)]{\includegraphics[width=0.5\textwidth]{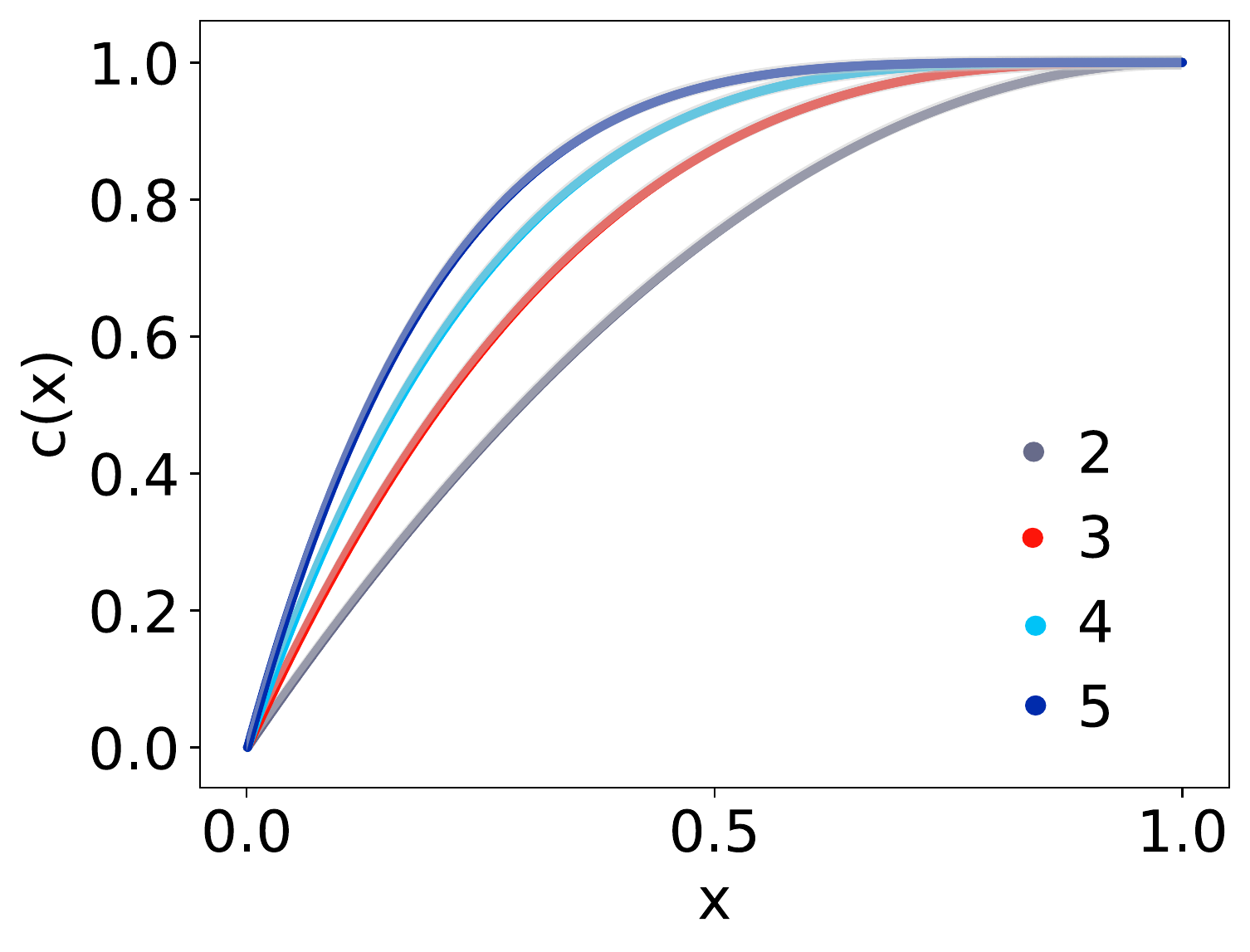}}
\caption{Combinatorial Approximation. The coloured lines show (a) the probability, and (b) the cummulative probability of the combinatorial summatory. The shaded area corresponds to the continuous approximation. The variable $x$ is the continuous equivalent of the discrete $i$ which allows us to associate each term in the summatory to a rational number in $[0,1]$.}
\label{fig:urhben}
\end{figure}

We provide here an example of the hyperlink counting process. Let us suppose that we have a system with $N=5$ and $g=3$, meaning that we could create a hypergraph of a potential number of $C^5_3=10$ hyperlinks, given by
\begin{equation*}
\mathcal{L}=\{123,124,125,134,135,145,234,235,245,345\}
\end{equation*}
The mechanism to a assign a random natural number in $[1,C^N_G]$ to each hyperlink is to make use of the star decomposition. Following with our example, we have $C^5_3=C^4_2 +C^3_2+ C^2_2$, which enables us to make a first partition on the g-star hypergraphs: $\{123,124,125,134,135,145\}$, $\{234,235,245\}$ and $\{345\}$. We can recursively apply our procedure, and make another partition in the subgraphs that we have here, for instance $\{123,124,125\}$, $\{134,135\}$ and $\{145\}$ conform the first set of the previous step.
These lists of combinatorial numbers may be approximated with a continuous function for better handling them. As discussed in the methods section, we propose to use $c_i \sim 1-(1-x)^g$ as an approximation for the cumulative probability, which gives $d_i \propto g(1-x)^{g-1}$ for the ordinary probability. In Figure \ref{fig:urhben} we show the convergence between the continuous approximation and the real discrete function. These formulas are crucial for carrying out the PRH algorithm.

\subsubsection*{Replicator Dynamics for URH}
Let us introduce now the replicator model for predicting the dynamics of the system. This technique is based on the indistinguishability between different nodes, and therefore it should describe the system as long as the hypergraph is connected and the hyperlinks are uniformly distributed among the nodes. We are working with URH, and therefore the more connected the system is, the better it fulfils the uniform requirement, and the closer it gets to the replicator description. 
 
In this simplified analysis we will describe the system in terms the fraction of defectors $x_D$ and the fraction of cooperators $x_C$. We compute the average normalized payoff of defectors and cooperators $\pi_D$ and $\pi_C$, as a function of their fractions, $x_D$ and $x_C$, in a hyperlink of order $g$. In order to do so, we sum the payoff of all the possible configurations of $g-1$ nodes. The sum is taken over $g-1$ because one of the nodes is already occupied by a defector, in $\pi_D$, or a cooperator in $\pi_C$. The payoff is calculated with the reduced synergy factor $r$.
\begin{eqnarray}
\nonumber \pi_D&=&\sum_{i=0}^{g-1} \left( \begin{array}{c} g-1 \\ i \end{array} \right) x^{g-1-i}_D x^{i}_C i r  \\ \pi_C&=&\sum_{i=0}^{g-1} \left( \begin{array}{c} g-1 \\  i \end{array} \right) x^{g-1-i}_D x^{i}_C ((i+1)r -1) 
\label{pay}
\end{eqnarray}    
The total average payoffs, are obtained by multiplying the ones of Equation \eqref{pay} with the average hyperdegree. But, we also normalize the total payoff with the hyperdegree, so the terms that go into the dynamical equation are indeed $\pi_D$ and $\pi_C$. One can observe that the expression for $\pi_D -\pi_C$ does not contain an explicit dependence on $g$, and therefore the same formula is valid for any order.
\begin{eqnarray}
\nonumber
\pi_D - \pi_C &=& \sum_{i=0}^{g-1} \left( \begin{array}{c} g-1 \\ i \end{array} \right) x^{g-1-i}_D x^{i}_C   [i r -  ((i+1)r -1)] \\ \nonumber &=&  (1-r) \sum_{i=0}^{g-1} \left( \begin{array}{c} g-1 \\ i \end{array} \right) x^{g-1-i}_D x^{i}_C  \\ &=& (1-r) (x_D + x_C)^{g-1}  = 1-r
\label{invariant0}
\end{eqnarray}

We now compute $\Delta$, the maximal payoff difference to be used in the replicators update rule. The $\pi^{+}$ and $\pi^{-}$ indicate the maximal and minimal payoff values from all the possible configuration of strategies.  
\begin{equation*}
\Delta \equiv \left\{\begin{array}{l} \pi^{+}_{D} -\pi^{-}_{C} \hspace{0.2cm}
\textrm{if} \hspace{0.2cm} \pi^{+}_{D} - \pi^{-}_{C} > \pi^{+}_{C} - \pi^{-}_{D}  \\ \pi^{+}_{C} - \pi^{-}_{D}\hspace{0.2cm}
\textrm{if} \hspace{0.2cm}  \pi^{+}_{D} - \pi^{-}_{C} < \pi^{+}_{C} - \pi^{-}_{D} \end{array}\right.
\end{equation*}
$\Delta$ is only expressed for comparisons of payoffs of different strategies, since the strategies are copied from the neighbours, and therefore the terms $\pi^{+}_{i} -\pi^{-}_{i}$, where two nodes share the same strategy, are irrelevant. For any order $g$, $\Delta$ is given by
\begin{equation}
\Delta = \left\{\begin{array}{l} r(g-2)+1 \hspace{0.2cm} \textrm{if} \hspace{0.2cm} r < 1  \\ 
gr-1 \hspace{0.2cm} \textrm{if} \hspace{0.2cm}  r > 1 \end{array}\right.
\label{delta}
\end{equation}
We now work with $Q$, the normalized payoff difference, $Q\equiv (\pi_D - \pi_C) / \Delta$.

The system equation is derived as the sum of all the channels via which the strategy of a node can change. These channels correspond to the different configurations of strategies of hyperlinks of order $g$, weighted with the probabilities of the nodes having a particular strategy. For each configuration, one has to compute the probability of having a defector-cooperator pair, $w_D w_C / g (g-1)$ and the probability that this pair results in a strategy change. Notice that the values of $w_D$ and $w_C$, the number of defectors and cooperators in the group, are precisely the exponents of $x_C$ and $x_D$. We employ the replicator update, which in this context means that the flip probability is computed as $-Q \theta(\pi_C - \pi_D)$ when a defector changes into a cooperator and as $Q \theta(\pi_D -\pi_C)$ in the opposite case. This sign corresponds to the calculation of the fraction of defectors, the contrary holds true for the fraction of cooperators.

\begin{eqnarray*}
\partial_t x_D &=& \sum^{G-2}_{i=0} \left( \begin{array}{c} g \\ 1+i \end{array} \right) x^{g-1-i}_D x^{1+i}_C \frac{(g-1-i)(1+i)}{g(g-1)} \Big{[} - \big{(}-Q \theta(\pi_C - \pi_D)\big{)} +Q\theta(\pi_D -\pi_C) \Big{]} 
\\ &=& \sum^{G-2}_{i=0} \left( \begin{array}{c} g \\ 1+i \end{array} \right) x^{g-1-i}_D x^{1+i}_C \frac{Q (g-1-i) (1+i)}{g(g-1)} \\ &=& \sum^{G-2}_{i=0} Q x^{g-1-i}_D x^{1+i}_C  \frac{g! (g-1-i) (1+i)}{(g-1-i)! (1+i)! g(g-1)}  
\\ &=& \sum^{G-2}_{i=0} Q x^{g-1-i}_D x^{1+i}_C  \frac{(g-2)! }{(g-2-i)! i!} \\ &=& Q x_D x_C \sum^{G-2}_{i=0} \left( \begin{array}{c} g-2 \\ i \end{array} \right) x^{g-2-i}_D x^{i}_C =
Q x_D x_C (x_D + x_C)^{g-2} = Q x_D x_C
\end{eqnarray*} 

This is precisely the replicator equation which provides a very useful tool for describing the system since its structure does not explicitly depend on $g$, even if the evolution itself does, since $Q$ is a function of $g$.
\begin{eqnarray}
\nonumber
\partial_t x_D &=& Q x_D x_C \\
\partial_t x_C &=& -Q x_D x_C 
\label{dyn}
\end{eqnarray}

The stationary condition yields three possible outcomes, $x_D=0,x_C=0$ and $Q=0$. The first two are the trivial phases of the system, and the third one denotes the critical point, that corresponds to $r=1$. Notice that this condition is reduced to finding the zeros of Equation \eqref{invariant0}, and is therefore common for all systems, and independent of their order. This implies that for random uniform hypergraphs cooperators only emerge when $R>g$.

If the initial condition is given by a uniform distribution of strategies, $x_D=x_C=0.5$, the time evolution reads
\begin{eqnarray*}
x_D (t) &=& \frac{1}{1+e^{ Qt}} \\
x_C (t) &=& \frac{1}{1+e^{-Qt}}  
\end{eqnarray*}
Even if the relaxation time is infinite, we can get the time to arrive the neighbourhood of the asymptotic state by imposing the condition $x_i = 1/N$ or $x_i = (N-1)/N$ that yield
\begin{equation}
T=\frac{\ln(N-1)}{|Q|}
\label{time}
\end{equation}

\subsection*{Hyperdegree-Heterogeneous Random Hypergraphs}
\subsubsection*{PRH and SRH}
\begin{figure}[h]
\captionsetup[subfigure]{labelformat=empty}
\centering
\subfloat[(a)]{\includegraphics[width=0.25\textwidth]{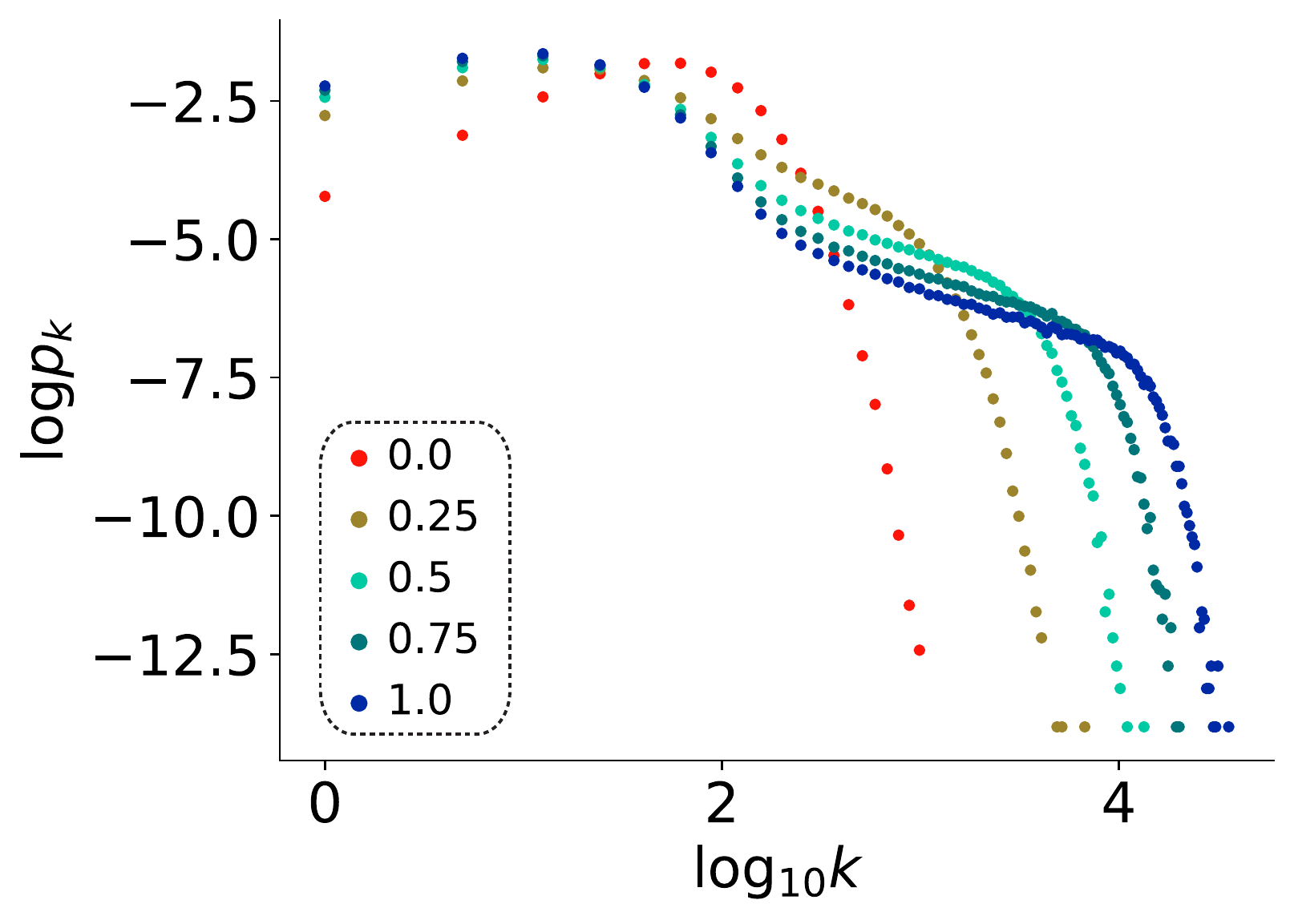}}
\subfloat[(b)]{\includegraphics[width=0.25\textwidth]{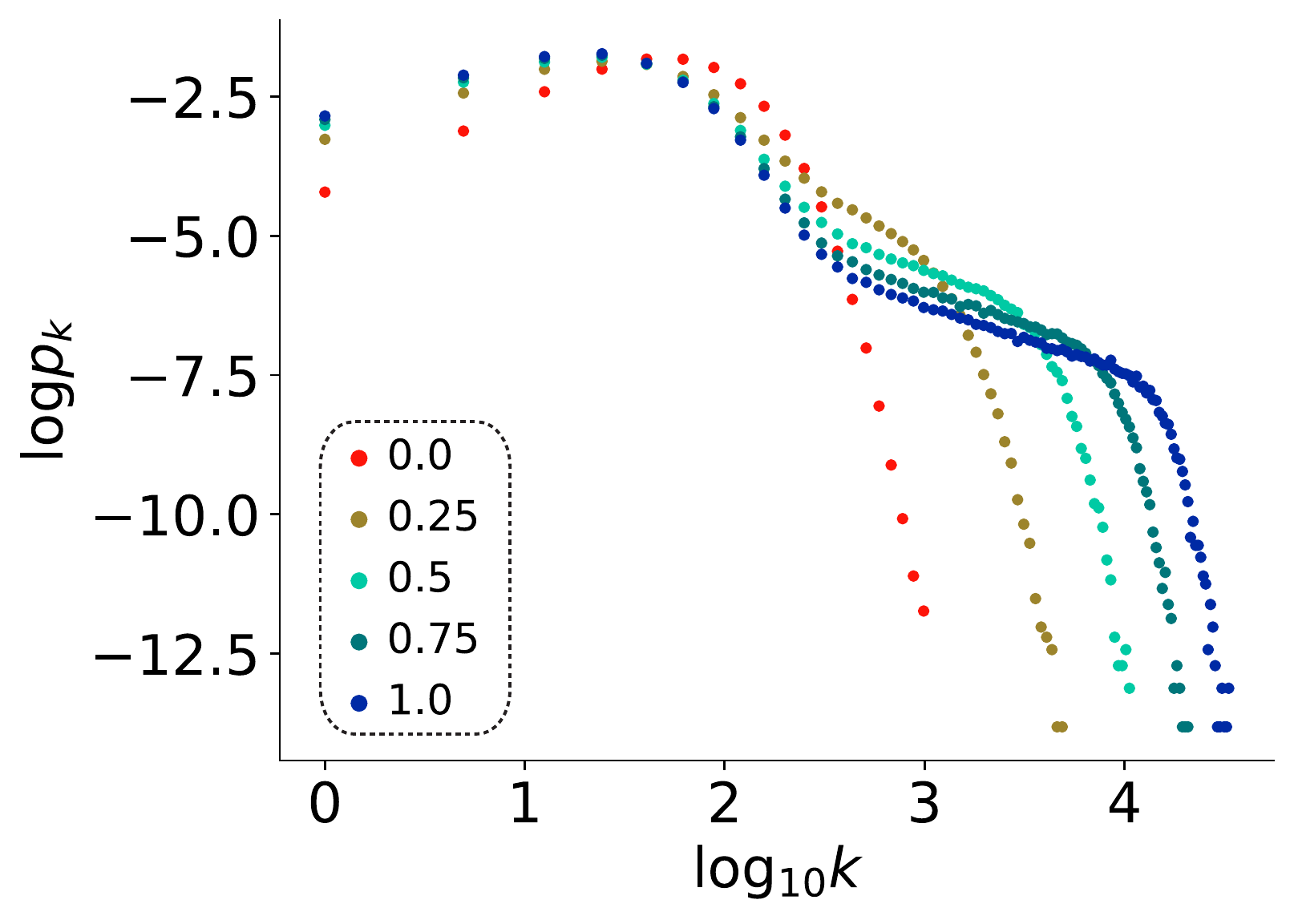}}
\subfloat[(c)]{\includegraphics[width=0.25\textwidth]{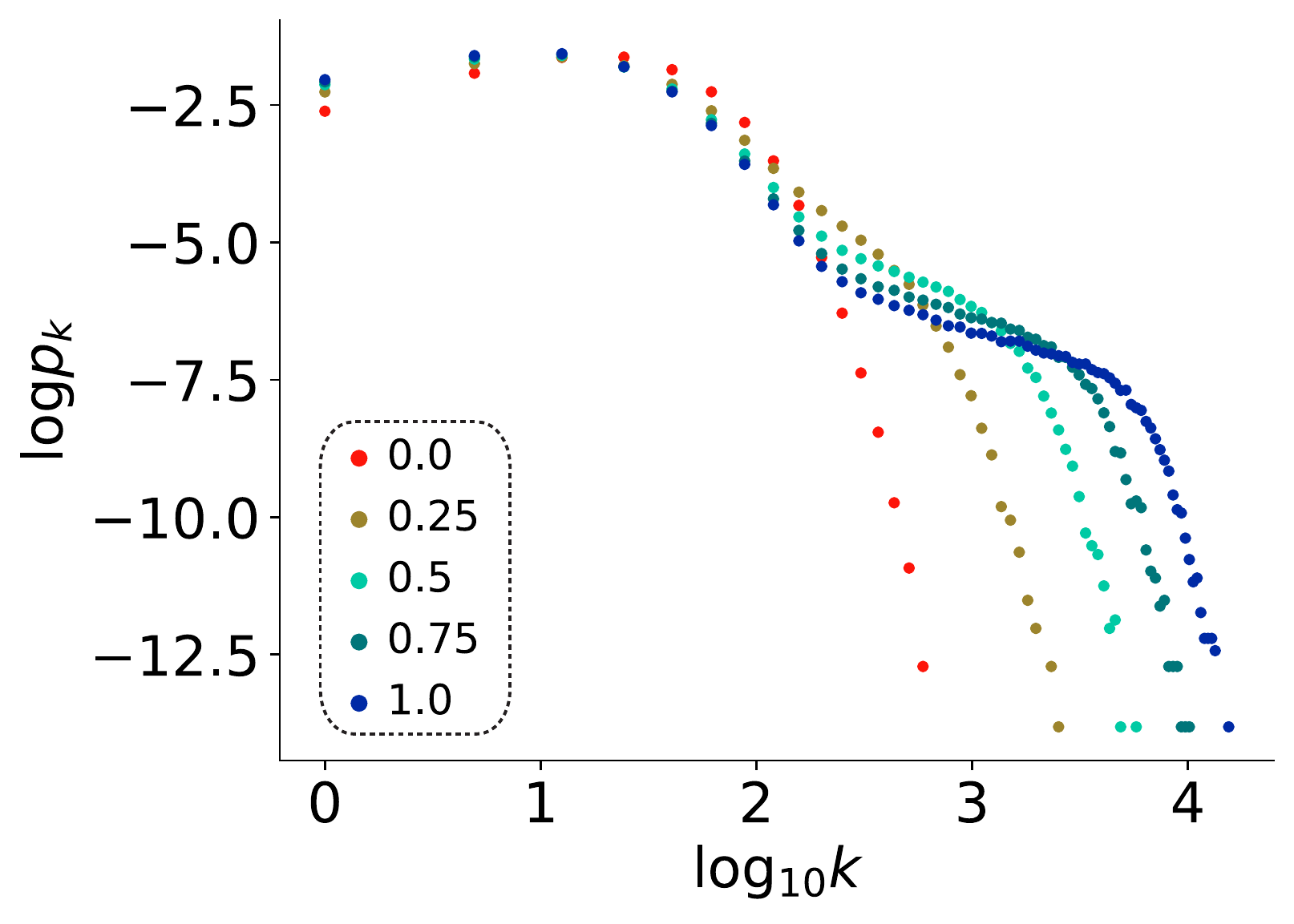}}
\subfloat[(d)]{\includegraphics[width=0.25\textwidth]{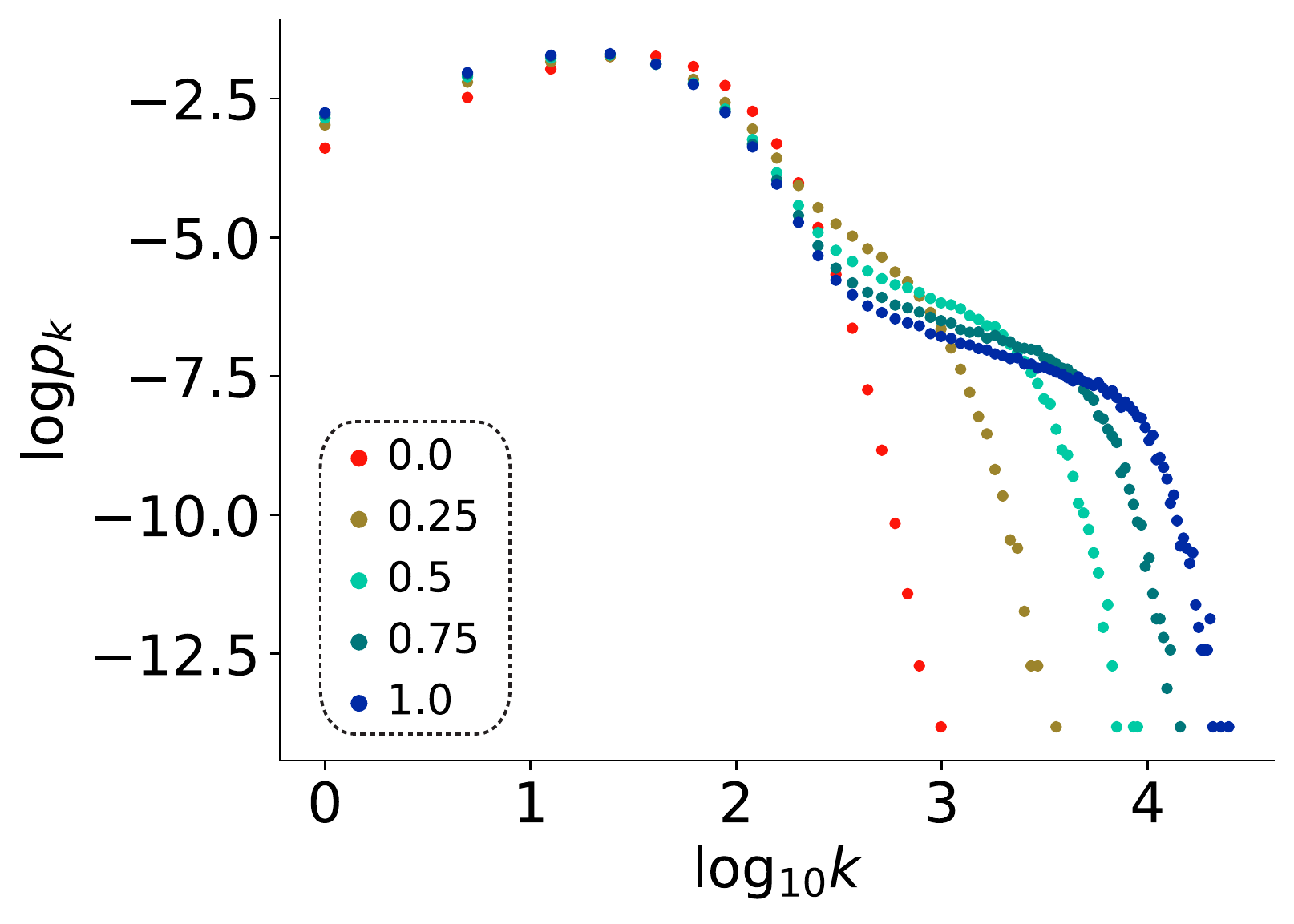}}

\subfloat[(e)]{\includegraphics[width=0.25\textwidth]{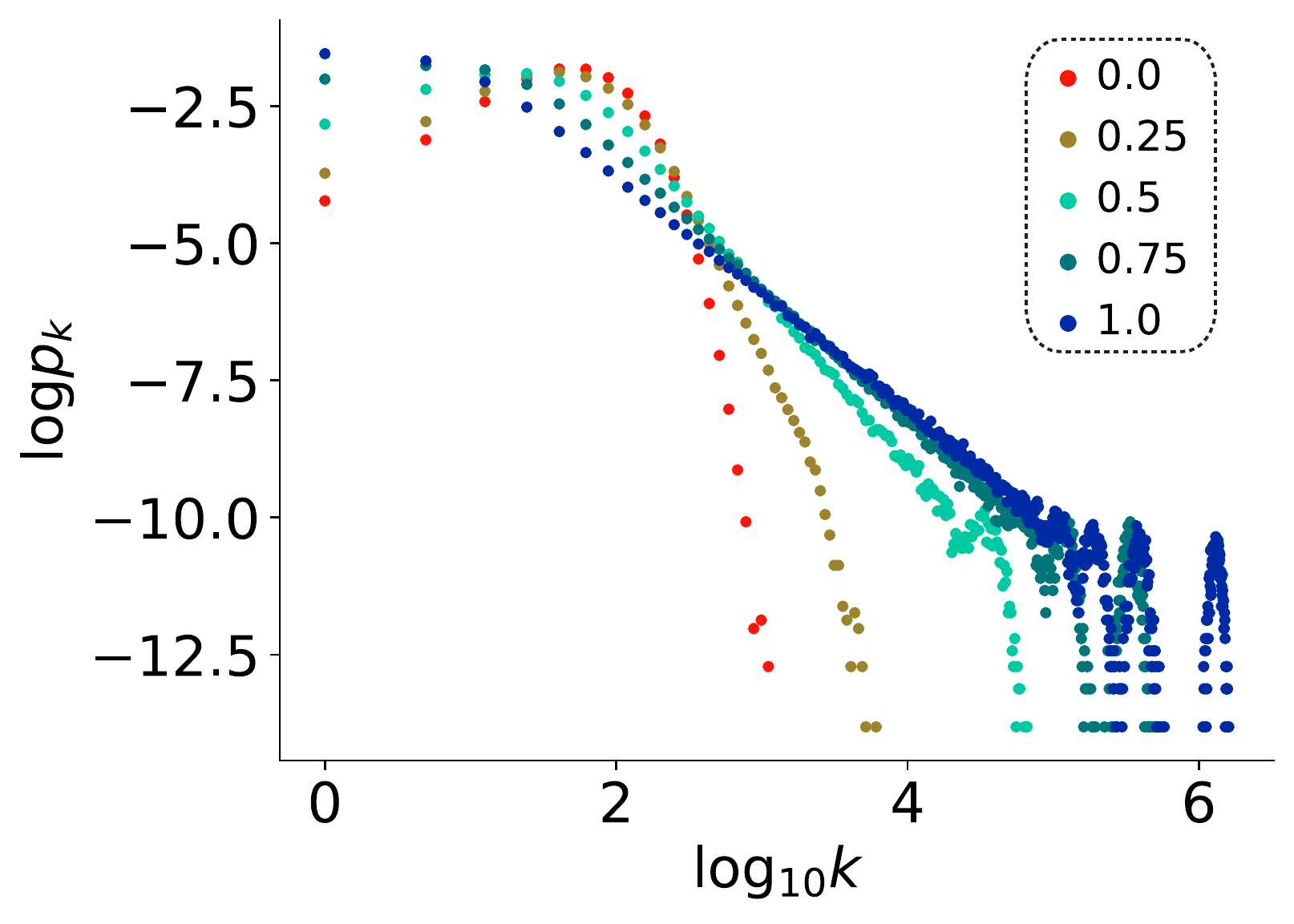}}
\subfloat[(f)]{\includegraphics[width=0.25\textwidth]{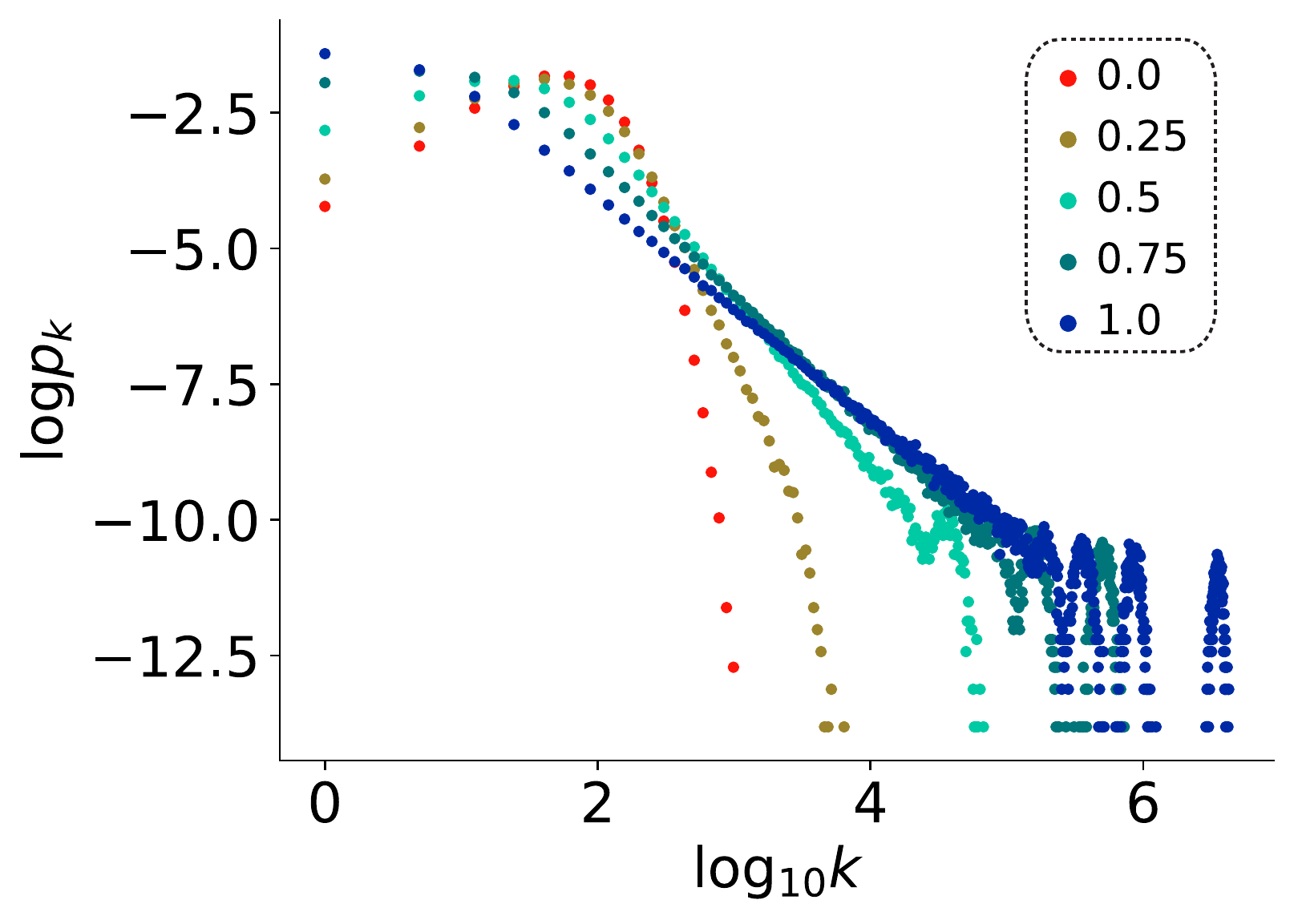}}
\subfloat[(g)]{\includegraphics[width=0.25\textwidth]{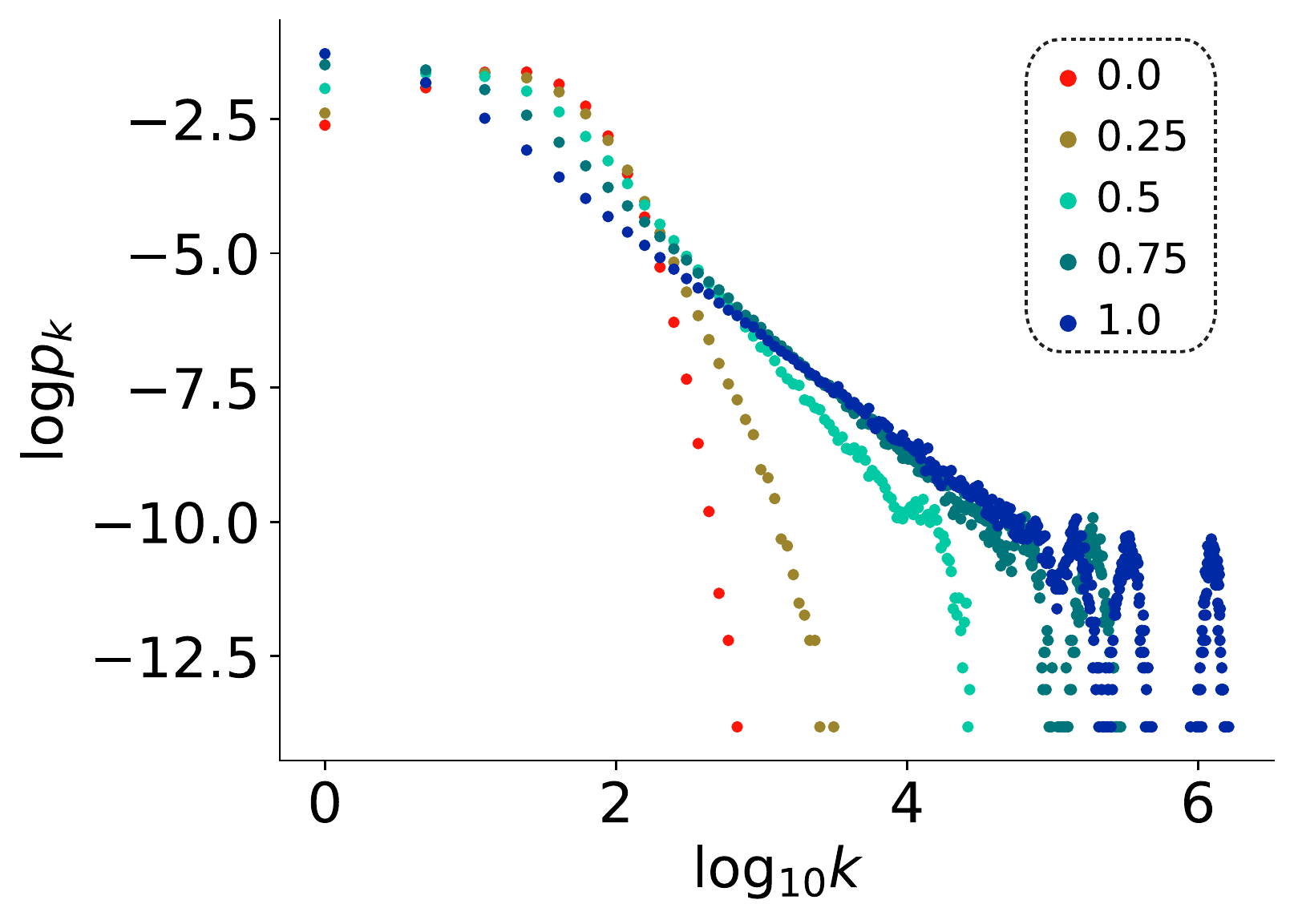}}
\subfloat[(h)]{\includegraphics[width=0.25\textwidth]{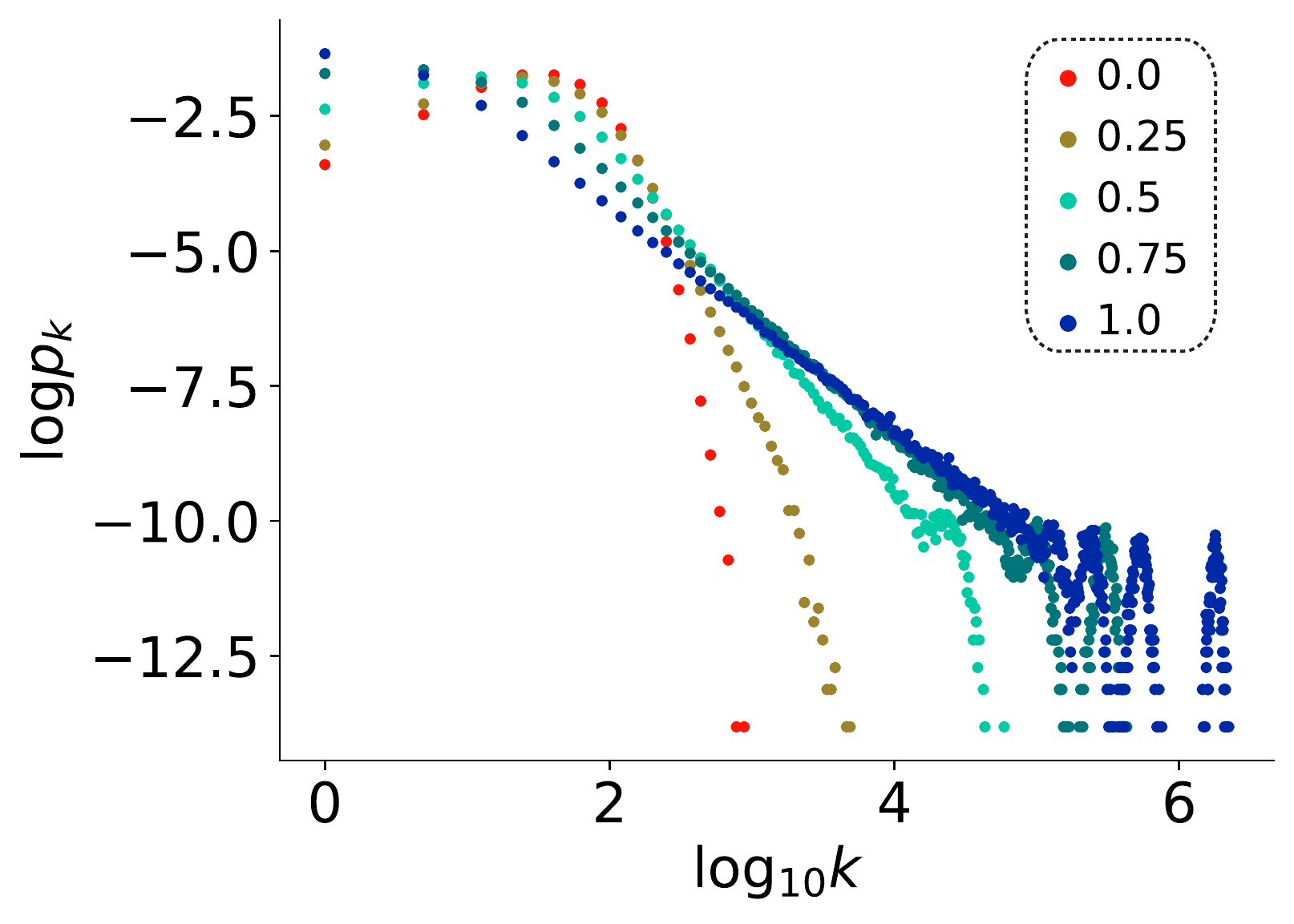}}
\caption{Hyperdegree-Heterogeneous Distributions. We show the value of $p_k$ the probability of having hyperdegree $k$ as a function of the value of the hyperdegree $k$ in a logarithmic scale. In the first and second row we plot the results for Power Random Hypergraphs and Scale-Free Random Hypergraphs respectively. Each of the columns is associated to $g=2,3,4,5$. In each of the subplots we plot the distributions obtained for $\mu=0,0.25,0.5,0.75,1$. The simulations have been performed over ensembles of Hypergraphs with $N=1000$ nodes and $L=L_c$ hyperlinks.}
\label{fig:rhben}
\end{figure}

In this subsection we report a statistical analysis about the average hyperdegree distribution obtained when creating the hypergraph using the PRH and SRH algorithms explained in the Methods section of the main text. We supply numerical evidence about the nature of the PRH and SRH, see Figure \ref{fig:rhben}. For the former the hyperdegree distribution has two clear phases, sharp decay and an intermediate flatter regime. This combination enables the existence of hubs with a maximum hyperdegree of $1.5$ orders of magnitude larger than in the uniform case. For the latter the hyperdegree distribution follows a power law, visualized as a straight line in the log-log scale. The hubs created using the SRH have a hyperdegree $3$ orders of magnitude larger than those in the uniform case.

\subsubsection*{Replicator Dynamics for Hyperdegree-Heterogeneous Hypergraphs}
In this subsection we provide theoretical arguments that clarify the results of the numerical simulations. More precisely, we are looking for the conditions that allow hypergraphs with a heterogeneous hyperdegree distribution to behave according to the dynamics derived for the uniform case. In order to do so, we are going to repeat all the steps in the formalization of the mean-field dynamics introducing an explicit dependence on the hyperdegree, such that the fraction of defectors and cooperators are expressed in terms of the corresponding fractions conditioned to the nodes hyperdegree
\begin{eqnarray}
\nonumber
x_D= \sum_{k \in \mathcal{K}} p(k) p(D|k) \\
x_C= \sum_{k \in \mathcal{K}} p(k) p(C|k)
\label{hetmf}
\end{eqnarray}
Here $\mathcal{K}$ denotes the set of all possible hyperdegrees and $k$ an element of that set.\\ 

We first notice that the value of $\Delta$ is the same as in the uniform case, as $\Delta$ is a function of the maximal and minimal possible payoffs, which are the same. We continue with the proposal of the dynamical equation. The time evolution of Equation \eqref{hetmf} can be expressed as a sum of each of the $p(D|k)$ terms, that for clarity we relabel as $x_{Dk}$. There are two possible transitions in which the fraction of defectors with hyperdegree $k$, $x_{Dk}$ can change: a cooperator with hyperdegree $k$ changes into a defector of hyperhyperdegree $k$, and a defector of hyperdegree $k$ changes into a cooperator of hyperdegree $k$. The dynamics accounts for these two contributions as follows: 
\begin{eqnarray*}
\frac{d}{dt} x_{Dk}&=& -x_{Dk} \sum_{k' \in \mathcal{K}} p(k'|k) p(C|k') \sum_{k'' \in \mathcal{K}^{G-2}} p(k''|kk') \sum_{x \in \mathcal{X}^{G-2}} p(x|k'') \times \\ &&\frac{(\pi_{Ck'}-\pi_{Dk})\theta(\pi_{Ck'}-\pi_{Dk})}{\Delta} \\
&&+x_{Ck} \sum_{k' \in \mathcal{K}} p(k'|k) p(D|k') \sum_{k'' \in \mathcal{K}^{G-2}} p(k''|kk') \sum_{x \in \mathcal{X}^{G-2}} p(x|k'') \times \\ &&\frac{(\pi_{Dk'}-\pi_{Ck})\theta(\pi_{Dk'}-\pi_{Ck})}{\Delta} \\
\end{eqnarray*}
The variable and first summatory in both terms account for the probability of having the defector and cooperator pair that will take part in the strategy update. In order to ensure that both are in the same hyperlink we introduce the hyperdegree-hyperdegree correlation to account for the probability that the hyperdegree of the player inspiring the strategy update is $k'$, given that the player changing is strategy has hyperdegree $k$. The next term corresponds to the remaining $g-2$ nodes of the hyperlink, and thus, $k''$ represents $G-2$ elements of $\mathcal{K}$. Again the hyperdegree-hyperdegree correlation provides the information about the conditional probability of having $g-2$ nodes with hyperdegrees $k''$ given that the remaining two nodes have hyperdegrees $k$ and $k'$. In the next term, the summation in $x$, $x$ accounts for $G-2$ elements of $\mathcal{X}$, where $\mathcal{X}$ is the set of all the possible strategies, either $D$ or $C$. The last term contains the comparison of the payoffs, that will eather increase or decrease the amount of defectors. Following the same strategy as in the uniform case, we will have a closer look at the payoff comparison. The $\Delta$ contribution is solved already, so we focus on the numerator. 
\begin{eqnarray*}
\pi_{Dk} &=& \sum_{k'' \in \mathcal{K}^{G-1}} p(k''|k) \sum_{x'' \in \mathcal{X}^{G-1}} p(x''|k'') (nr) \\
\pi_{Ck'} &=& \sum_{k'' \in \mathcal{K}^{G-1}} p(k''|k') \sum_{x'' \in \mathcal{X}^{G-1}} p(x''|k'') ((n+1)r-1)
\end{eqnarray*}
Here $n$ is the number of cooperators of each particular case of the summatory, which is of course a function of $x''$.\\

The important terms are $p(k''|k)$ and $p(k''|k')$, those that account for the hyperdegree-hyperdegree distribution, i.e., the probability that a node with a given hyperdegree $k$ or $k'$ is part of a hyperlink with nodes of hyperdegrees given by $k''$. For hypergraphs in which the neighbouring hyperdegree distribution $p(k''|k)$ is given by the total hyperdegree distribution $p(k'')$, this expression can be reduced to
\begin{eqnarray*}
\pi_{Dk} &=& \sum_{k'' \in \mathcal{K}^{G-1}} p(k'') \sum_{x'' \in \mathcal{X}^{G-1}} p(x''|k'') (nr) \\
\pi_{Ck'} &=& \sum_{k'' \in \mathcal{K}^{G-1}} p(k'') \sum_{x'' \in \mathcal{X}^{G-1}} p(x''|k'') ((n+1)r-1)
\end{eqnarray*}
and then
\begin{eqnarray*}
\pi_{Dk} - \pi_{Ck'} &=& \sum_{k'' \in \mathcal{K}^{G-1}} p(k'') \sum_{x'' \in \mathcal{X}^{G-1}} p(x''|k'') ((nr)-((n+1)r-1)) \\
&=& \sum_{k'' \in \mathcal{K}^{G-1}} p(k'') \sum_{x'' \in \mathcal{X}^{G-1}} p(x''|k'') (1-r) \\
&=& (1-r)
\end{eqnarray*}
This is indeed the same result as in the uniform case, $Q=\frac{1-r}{\Delta}$, an expression that we can now use in the main derivation to simplify the equation  
\begin{eqnarray*}
\frac{d}{dt} x_{Dk}&=& -x_{Dk} (-Q \theta(r-1)) \sum_{k' \in \mathcal{K}} p(k'|k) p(C|k') \sum_{k'' \in \mathcal{K}^{G-2}} p(k''|kk') \sum_{x \in \mathcal{X}^{G-2}} p(x|k'') \\
&&+x_{Ck} Q \theta(1-r) \sum_{k' \in \mathcal{K}} p(k'|k) p(D|k') \sum_{k'' \in \mathcal{K}^{G-2}} p(k''|kk') \sum_{x \in \mathcal{X}^{G-2}} p(x|k'') \\
&=& Q \theta(r-1) x_{Dk} x_C + Q \theta(1-r) x_{Ck} x_D 
\end{eqnarray*}
In the second step we have make use of the assumption that there are not hyperdegree-hyperdegree correlations, $p(k'|k)=p(k')$. We then insert this equation into Equation \eqref{hetmf} which yields 
\begin{eqnarray*}
\frac{d}{dt} x_D = Q \theta(r-1) x_D x_C + Q \theta(1-r) x_C x_D = Q x_D x_C
\end{eqnarray*}
The dynamics is equivalent to the one derived in the uniform case. The interpretation is the following: when the nodes are indistinguishable in the sense that their neighbourhoods are equivalent the mean-field dynamics is still valid even if the hypergraphs are heterogeneous.  

\subsection*{Order-Heterogeneous Random Hypergraphs}
\subsubsection*{Replicator Dynamics for Order-Heterogeneous Hypergraphs}
The first thing to notice is that the average payoff difference, is nothing but the sum of the average payoff differences for each of the orders, weighted with the $p^g$. The sum goes from $g^-$, the minimal value of $g$ to $g^+$, the maximal one.
\begin{equation*}
\pi_D - \pi_C=\sum^{g_+}_{g=g_-} p^{g} (1-r^{g}) 
\end{equation*}

We first derive the normalization factor $\Delta$, from Equation \eqref{delta}, via the explicit expression for the reduced synergy factor $r^g=\alpha g^{\beta -1}$. We consider $\alpha, \beta \ge 0$. The extreme point of $r^g$ at $\beta=1$ divides the analysis in two different intervals, $\beta<1$ and $\beta \ge 1$. 

In the first one, $\beta <1$, the maximal payoffs $\pi^{+}$ are obtained for the lowest order $g_{-}$, and the minimal payoffs $\pi^{-}$ for the highest order $g_{+}$.
\begin{equation*}
\pi^{+}_{D} = \alpha g^{\beta -1}_{-} (g_- -1), \hspace{0.5cm} \pi^{-}_{D} = 0, \hspace{0.25cm} \pi^{+}_{C} = \alpha g^{\beta}_{-} -1 , \hspace{0.25cm} \pi^{-}_{C} = \alpha g^{\beta -1}_{+} -1
\end{equation*}

Therefore, we get

\begin{equation*}
\Delta (\beta < 1) = \left\{\begin{array}{l} 
\alpha g^{\beta -1}_{-} (g_- -1) -\alpha g^{\beta -1}_+ +1  \hspace{0.2cm} \textrm{if} \hspace{0.2cm} \alpha \le \frac{2}{g^{\beta -1}_- + g^{\beta-1}_+}  \\ 
\alpha g^{\beta}_- -1 \hspace{0.2cm} \textrm{if} \hspace{0.2cm} \alpha > \frac{2}{g^{\beta -1}_- + g^{\beta-1}_+}
\end{array}\right.
\end{equation*}

In the second interval, $\beta \ge 1$, the maximal payoffs are obtained for the maximal orders, and the minimal payoffs for the minimal orders.
\begin{equation*}
\pi^{+}_{D} = \alpha g^{\beta -1}_{+} (g_+ -1), \hspace{0.5cm} \pi^{-}_{D} = 0, \hspace{0.25cm} \pi^{+}_{C} = \alpha g^{\beta}_{+} -1 , \hspace{0.25cm} \pi^{-}_{C} = \alpha g^{\beta -1}_{-} -1
\end{equation*}

which yields,

\begin{equation*}
\Delta (\beta \ge 1) = \left\{\begin{array}{l} 
\alpha g^{\beta -1}_{+} (g_+ -1) -\alpha g^{\beta -1}_- +1  \hspace{0.2cm} \textrm{if} \hspace{0.2cm} \alpha \le \frac{2}{g^{\beta -1}_- + g^{\beta-1}_+}  \\ 
\alpha g^{\beta}_+ -1 \hspace{0.2cm} \textrm{if} \hspace{0.2cm} \alpha > \frac{2}{g^{\beta -1}_- + g^{\beta-1}_+}
\end{array}\right.
\end{equation*}

With $\pi_D-\pi_C$ and $\Delta$, we can obtain $Q$, as in the uniform case. And in terms of $Q$ the differential equation for the time evolution is exactly Equation \eqref{dyn}. Therefore, the relaxation time is again given by Equation \eqref{time}, and the critical point is again obtained by making $\pi_D-\pi_C=0$, which yields
\begin{equation*}
\alpha_c=\frac{1}{\sum^{g_+}_{g=g_-} p^g g^{\beta -1} }
\end{equation*}

\subsubsection*{Hypergraphs describing real-world collaborations}

\begin{figure}[h]
\captionsetup[subfigure]{labelformat=empty}
\centering
\subfloat[]{\includegraphics[width=\textwidth]{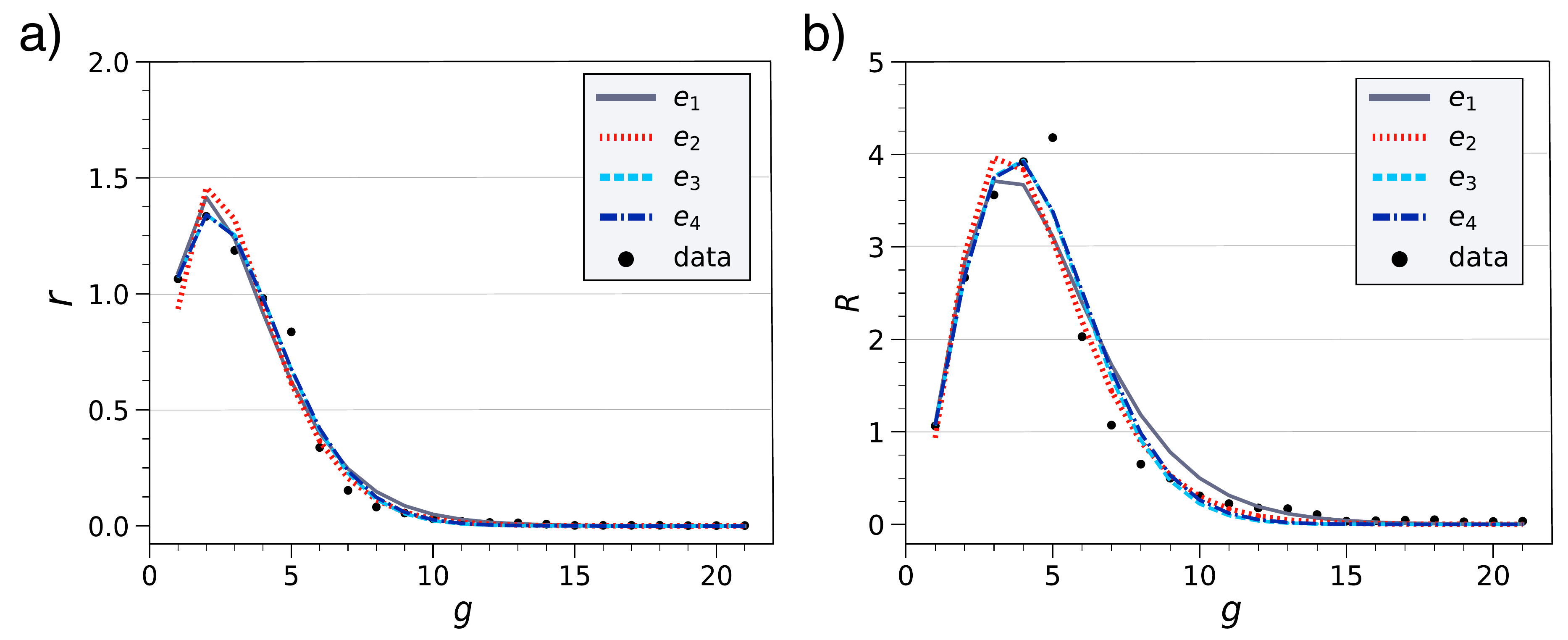}}
\caption{Synergy factors of the Internet Engineering Task Force dataset. We extract the synergy factor as a function of the
  group size for a bibliographic dataset in technology that contains the
  number of publications as a function of the number of authors
  $g$. We infer the distribution in the order of the hyperdegrees from the
  original dataset by assuming that the hyperlinks are evenly distributed amongst the nodes. We then
  impose the critical point condition and extract the value of the
  synergy factor under the hypothesis that $r(g)$ is proportional to
  ${\bf p}$. Finally, the synergy factors  $r(g)$ in (a) and 
$R(g)$ in (b) are factorized into two
  analytical expressions that respectively account for the benefits of large cooperations, and the costs associated to saturation effects.}
\label{dat}
\end{figure}

\paragraph*{IETF Dataset.}
In this last section, we provide an additional practical example of the procedure for extracting the synergy factor from real data, based on the publication records of the Internet Engineering Task Force (IETF). This is a large open international community of network designers, operators, vendors, and researchers concerned with the evolution and the smooth operation of the Internet. Similarly to the APS journals, the system is described by a hypergraph where authors who participated together in a project are part of the same hyperlink. For the IETF dataset, we have 34280 articles. In Figure \ref{dat} we report the histogram of the number of articles with a given number of authors. The normalized version of this histogram is precisely the hyperlink distribution as a function of the order of the hyperlink. Next, we retrieve the hyperdegree-order distribution by assuming that the hyperlinks are equally distributed amongst the nodes, meaning that for each order $g$ the hyperdegree can be deduced from the number of hyperlinks as $k^g=gL^g/N$. We assume that the reduced synergy factor is proportional to the hyperdegree distribution. This assumption means that we need an additional parameter to provide the full expression of the reduced synergy factor (and not the normalized one). This parameter can be obtained by introducing the critical point condition $\sum^{g=g^+}_{g=g^-} p^g (1-r(g))=0$. Notice that $p^g$ is the $g$th component of the hyperdegree, and that $r(g)$ is the synergy factor we are looking for, which can be expressed
as $z p^g$ under the aforementioned assumption. This provides a simple equation to obtain $z$, $z \sum (p^g)^2 =1$, which completes the procedure for extracting $r(g)$.

For this particular dataset, for completeness we fit the experimental data with different analytical expressions. We numerically
find those parameters that minimize the distance with either $R(g)$ or $r(g)$. To validate the fit, the measure
we have selected is the distance between the normalized points of the analytical and the
experimental synergy factor, which is bounded to the $[0,1]$ interval. Therefore, values close
to $0$ account for a good estimation of the data, otherwise the opposite is true. To extract the parameters
we have performed a search process by limiting the possible values to meaningful regions of the parameter space, and
by discretizing this region into a finite set values. Then, we iterated over all the sets, one for each
parameter, and calculated the normalized distance between the analytical expression and the
experimental one for each iteration. The best results that we have obtained in Figure \ref{dat} are
given by
\begin{eqnarray*}
&& e_1: r(g)=\alpha g^{\beta} e^{-\gamma (g-1)} \hspace{0.6cm} (\beta,\gamma)=(1.382,0.696) \hspace{0.6cm} (d_r,d_R)=(0.054,0.090) \\
&& e_2: R(g)=\alpha g^{\beta} e^{-\gamma (g-1)} \hspace{0.6cm} (\beta,\gamma)=(2.89,0.865) \hspace{0.6cm} (d_r,d_R)=(0.065,0.082) \\
&& e_3: r(g)=\alpha g^{\beta} e^{-\gamma (g-1)^{\delta}} \hspace{0.6cm} (\beta,\gamma,\delta)=(0.5,0.12,1.7) \hspace{0.6cm} (d_r,d_R)=(0.039,0.079) \\
&& e_4: R(g)=\alpha g^{\beta} e^{-\gamma (g-1)^{\delta}} \hspace{0.6cm} (\beta,\gamma,\delta)=(1.51,0.13,1.65) \hspace{0.6cm} (d_r,d_R)=(0.041,0.077) \\
\end{eqnarray*}
Here $e_i$ is associated to the expressions as reported in the figure, $\alpha$ is obtained analytically by applying the critical point condition
and $d_r$ and $d_R$ the distance between the approximation and the points inferred from the
data. Notice that the $(e_1,e_2)$ and $(e_3,e_4)$ pairs of equations have the same structure, but their
parameters have been obtained by minimizing the distance with $r(g)$ and $R(g)$, respectively.
As before, $\beta$ and $\gamma$ are the benefits and costs fit parameter, that respectively grow as a power law and decrease
exponentially.

\paragraph*{APS Dataset.} Here we provide some additional details about the dataset of scientific collaborations from publications in the journals of the American Physical Society described in the main text. In Table 1 we report the main properties of the hypergraph describing each journal.\\
We clarify here the optimization method for $\beta$ and $\gamma$. Once we have selected
\begin{equation}
f(g,\alpha,\beta,\gamma)=\alpha g^{\beta} e^{-\gamma (g-1)}
\label{last}
\end{equation}
as the base equation to fit the data, we run a first coarse-grained analysis on the possible minimal and maximal limits of $\beta$ and $\gamma$. These limits account for the values outside which the approximation does not resemble the data. We then discretize the domain of $\beta$ and $\gamma$, and carry out a brute-force optimization, namely, we evaluate a distance $\sigma(r,f)$ for all the possible combinations of $\beta$ and $\gamma$ and select those with the smallest value. The function $\sigma(r,f)$ is bounded to the $[0,1]$ interval, and measures the distance between the normalized synergy factor $r$ and the normalized approximation $f$, calculated as $\sigma=1-\sum_g \min(r(g),f(g))$.\\

\begin{figure}[h]
\captionsetup[subfigure]{labelformat=empty}
\centering
\subfloat[]{\includegraphics[width=0.5\textwidth]{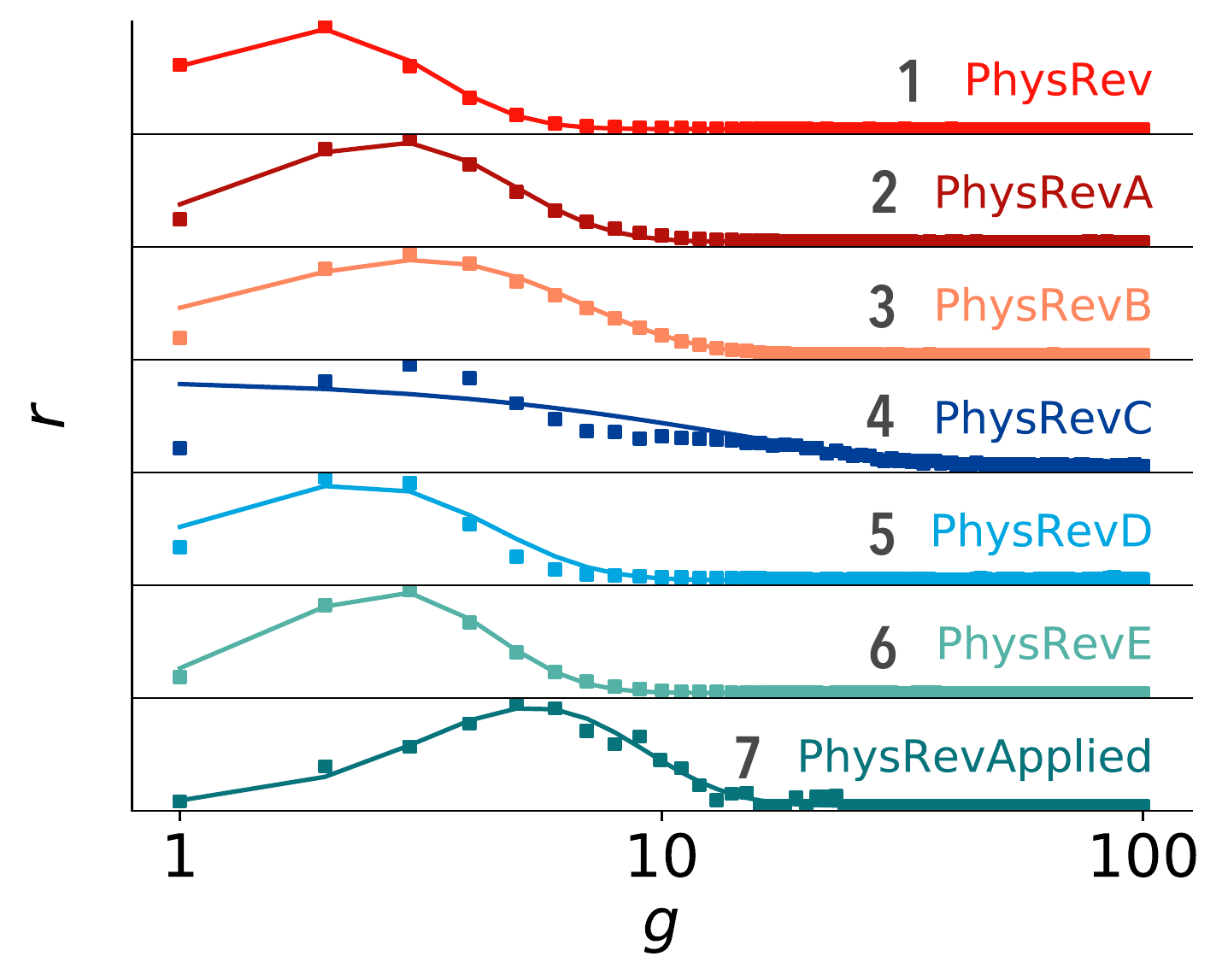}}
\subfloat[]{\includegraphics[width=0.5\textwidth]{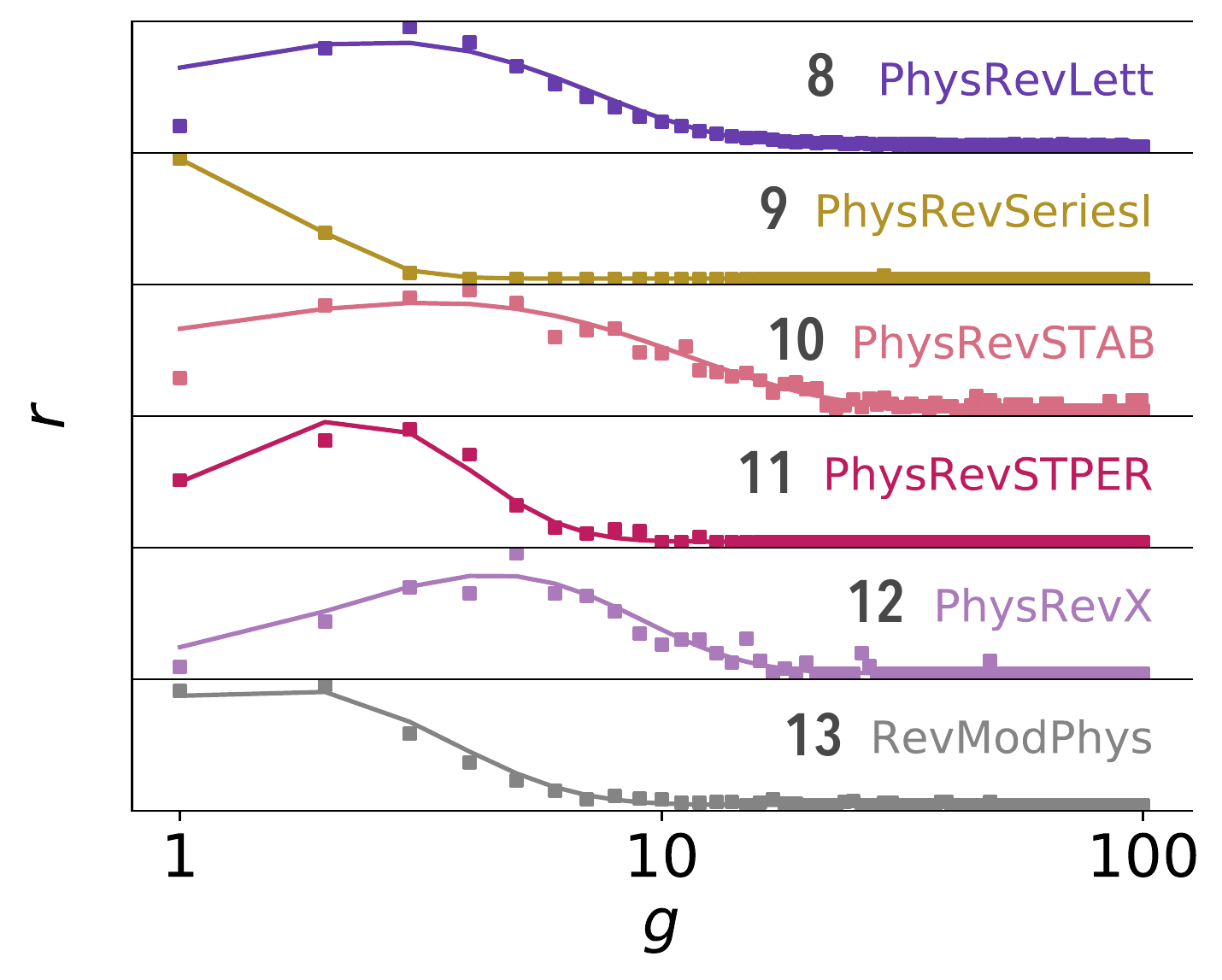}}
\caption{Synergy factors of the American Physical Society dataset. Value of the reduced synergy factor $r(g)$ for all the values of the group size $g$ available in the dataset. These plots are an extended version of Fig.4 shown in the main manuscript.}
\label{fig:aps}
\end{figure}

We also provide here the benefit and cost parameters associated to each journal, as well as the normalized error between the maximum synergy factor extracted directly from the data and from the analytical estimate.
In Figure \ref{fig:aps} we show the experimental and approximated reduced synergy factors $r(g)$ as a function of all hyperlink orders $g$ contained in the publications dataset. This figure is an extended version of Fig.4(a) in the main dataset.

\begin{table}[h!] 
\centering
\begin{tabular}{*{7}{|c}|}
\hline
Journal & $L$ & $<g>$ & $g(\max{r})$ & $\beta$ & $\gamma$ & $d_r$ \\ 
\hline
PhysRev & 47313  & 1.95 & 2 & 2.936 & 1.573 & 0.033 \\
\hline
PhysRevA & 70502 & 3.07 & 3 & 2.679 & 0.986 & 0.067 \\
\hline
PhysRevB & 171268 & 3.75 & 3 & 1.531 & 0.49 & 0.05 \\
\hline
PhysRevC & 36290 & 5.98 & 3 & 0.02 & 0.075 & 0.146 \\
\hline
PhysRevD & 74715 & 3.02 & 2 & 2.178 & 0.941 & 0.206 \\
\hline
PhysRevE & 50988 & 2.93 & 3 & 3.84 & 1.41 & 0.048 \\
\hline
PhysRevApplied & 327 & 5.39 & 5 & 3.356 & 0.62 & 0.09 \\
\hline
PhysRevLett & 113674 & 4.57 & 3 & 0.848 & 0.33 & 0.175 \\
\hline
PhysRevSeriesI & 1240 & 1.21 & 1 & 2.691 & 2.831 & 0.019 \\
\hline
PhysRevSTAB & 2393 & 5.52 & 4 & 0.566 & 0.173 & 0.127 \\
\hline
PhysRevSTPER & 484 & 2.42 & 3 & 2.75 & 1.21 & 0.078 \\
\hline
PhysRevX & 611 & 5.28 & 5 & 1.85 & 0.416 & 0.127 \\
\hline
RevModPhys & 3153 & 2.05 & 2 & 1.19 & 0.79 & 0.112 \\
\hline
\end{tabular}
\caption*{Table 1: American Physical Society Dataset. For each journal we report the total number of hyperlinks $L$, the average hyperlink order $<g>$, the order associated to the maximal synergy factor $g(\max r)$, the benefit parameter $\beta$ and the cost parameter $\gamma$ (from the analytical approximation of $r(g)$), and the normalized distance between the reduced synergy factor extracted from the data empirically and after the fit, that we indicate as $d_r$.}
\label{table:aps}
\end{table}

\clearpage

\clearpage

\end{document}